\documentclass{aa}
\usepackage{CJKutf8}
\usepackage{graphicx}
\usepackage{txfonts}
\usepackage{color}

\begin{document}

\title{Rapid formation of massive planetary cores in a pressure bump}
\author{
Tommy Chi Ho Lau
\begin{CJK*}{UTF8}{bsmi}(劉智昊)\end{CJK*}
\inst{\ref{inst1}}
\and
Joanna Dr{\c a}{\.z}kowska\inst{\ref{inst1},\ref{inst2}}
\and
Sebastian M. Stammler\inst{\ref{inst1}}
\and
Tilman Birnstiel\inst{\ref{inst1},\ref{inst3}}
\and
Cornelis P. Dullemond\inst{\ref{inst4}}
}

\institute{
University Observatory, Faculty of Physics, Ludwig-Maximilians-Universität München, Scheinerstr. 1, 81679 Munich, Germany
\label{inst1}
\and
Max Planck Institute for Solar System Research, Justus-von-Liebig-Weg 3, 37077 Göttingen, Germany
\label{inst2}
\and
Exzellenzcluster ORIGINS, Boltzmannstr. 2, D-85748 Garching, Germany
\label{inst3}
\and
Institut für Theoretische Astrophysik, Zentrum für Astronomie, Heidelberg University, Albert Ueberle Str. 2, 69120 Heidelberg, Germany
\label{inst4}
}

\date{Received 1 September 2022; accepted 8 November 2022}

\abstract
{Models of planetary core growth by either planetesimal or pebble accretion are traditionally disconnected from the models of dust evolution and formation of the first gravitationally bound planetesimals. State-of-the-art models typically start with massive planetary cores already present.}
{We aim to study the formation and growth of planetary cores in a pressure bump, motivated by the annular structures observed in protoplanetary disks, starting with submicron-sized dust grains.}
{We connect the models of dust coagulation and drift, planetesimal formation in the streaming instability, gravitational interactions between planetesimals, pebble accretion, and planet migration into one uniform framework.}
{We find that planetesimals forming early at the massive end of the size distribution grow quickly, predominantly by pebble accretion. These few massive bodies grow on timescales of $\sim$100 000 years and stir the planetesimals that form later, preventing the emergence of further planetary cores. Additionally, a migration trap occurs, allowing for retention of the growing cores. }
{Pressure bumps are favourable locations for the emergence and rapid growth of planetary cores by pebble accretion as the dust density and grain size are increased and the pebble accretion onset mass is reduced compared to a smooth-disc model. }

\maketitle
%-------------------------------------------------------------------

\section{Introduction}
Recent high-resolution interferometry observations by the Atacama Large Millimeter/submillimeter Array (ALMA) revealed that substructure may be common in protoplanetary discs, although we are still limited to the largest and thus brightest ones. Nevertheless, there is increasing evidence from comparison of the observational data of disc
populations and theoretical models that disc substructures must be common even in unresolved discs \citep{Toci2021,ZormpasApostolos2022}. Multiple surveys, such as those of \cite{Andrews2018}, \cite{Long2018}, and \cite{Cieza2020}, have shown that most of the substructures are presented in the form of axisymmetric rings. Through detailed analysis, \cite{Dullemond2018} found that dust trapping in a pressure bump is consistent with the rings sampled from the DSHARP survey. Kinematic studies with ALMA by \cite{Teague2018a,Teague2018} further showed that the dust rings coincide with the local pressure maxima in the analysed discs. However, the origin of such pressure bumps remains uncertain, where the possible causes include disc--planet interactions due to an existing massive planet \citep{Rice2006,PinillaP.2012,Dipierro2015,Dong2017}, sublimation \citep{Saito2011}, and instabilities \citep{Takahashi2014,FlockM.2015,Loren-Aguilar2015,PinillaPaola2016,DullemondC.P.2018}.

Despite the uncertainty as to the cause, the dust-trapping pressure bump is likely a favourable environment for the formation and growth of planetesimals towards massive planetary cores \citep{Morbidelli2020,GuileraOctavioMiguel2020,Chambers2021,Andama2022}. The locally enriched dust-to-gas ratio could trigger streaming instability \citep{Youdin2007,Johansen2007,Johansen2009,Bai2010}, which is the prevailing pathway to overcome the `metre-size barrier' of dust growth \citep{Weidenschilling1977,GuettlerC.2010,ZsomA.2010} to form planetesimals of on the order of 100 km in diameter \citep{JohansenA.2012,Johansen2015,Simon2016,Simon2017}. More recently, \cite{Stammler2019} suggested that planetesimal formation by streaming instability, where the midplane dust-to-gas ratio is regulated, can explain the similarity in the optical depths of the DSHARP rings studied by \cite{Dullemond2018}.  Furthermore, in the hydrodynamical simulations with self gravity by \cite{Carrera2021}, a small pressure bump can already trigger planetesimal formation by streaming instability with centimetre(cm)-sized grains, although this may not be applicable to the case with millimetre(mm)-sized dust \citep{Carrera2022}.

In the classical model, planetesimal accretion has been shown to quickly enter a stage of oligarchic growth with direct-$N$ body simulations \citep{Kokubo2000}. A massive planetary core is unlikely to form and accrete gas within the typical lifetime of protoplanetary discs -- particularly at wide orbits -- to form a cold Jupiter with the minimum mass solar nebula, while multiple works \citep[e.g.][]{FortierA.2007,FortierA.2009,GuileraO.M.2010} have shown that planetesimal accretion can be efficient in a significantly more massive disc. However, planetesimals that are large enough to gravitationally deflect dust from the gas streamline and have a sufficient encounter time can accrete a significant fraction of the drifting dust or `pebbles'. This emerged as a mechanism for efficient planetesimal growth often named `pebble accretion' (\citealp{OrmelC.W.2010,LambrechtsM.2012,GuillotTristan2014}; \citealp[see][for review]{Johansen2017,Ormel2017}). In a pressure bump, pebbles are trapped and the locally enhanced dust surface density also provides an elevated level of pebble flux compared to that in a smooth disc. The impeded drifting speed of the pebbles also lengthens the encounter time with the planetesimal, particularly in the outer disc where the pebble-carrying headwind is faster. Both of these factors increase the rate of planetesimal growth by pebble accretion inside a pressure bump.

As planetesimals grow into more massive embryos, the gravitational torque exerted by the disc becomes important. For low-mass planets that induce small perturbations in the disc, the disc--planet interaction lies in the type-I (or low-mass) regime (\citealp{Goldreich1979,Artymowicz1993,Tanaka2002,Tanaka2004}; \citealp[see][for review]{Kley2012}). As shown in the recent $N$-body planet-formation models (e.g. \cite{MatsumuraSoko2017,BitschBertram2019,LiuBeibei2019}), this poses a challenge: embryos formed in the outer Solar System should be stopped from entering the terrestrial planet region as they are needed in the outer Solar system to form cold giant planets. However, \cite{Coleman2016} showed that radial substructures in the disc can trap embryos at the outer edges and allow the formation of cold Jupiters. This suggests that pressure bumps are not only favourable to the growth of planetesimals but are also capable of retaining the massive planetary cores produced.

In this work, we present a numerical model of the formation and evolution of planetesimals in an axisymmetric pressure bump of a protoplanetary disc. We coupled the dust and gas evolution code \texttt{DustPy} \citep{Stammler2022} with the parallelized symplectic $N$-body integrator \texttt{SyMBAp} (Lau \& Lee in prep.) with modifications to include gas drag, type-I damping and migration, and pebble accretion according to the disc model. As the disc evolves and accumulates dust at the pressure bump, a fraction of the dust is converted into planetesimals according to the condition for streaming instability. These planetesimals are then realised as $N$-body particles and evolve through gravitational interactions as well as the additional processes mentioned above. The details of our models and their initial conditions are presented in Section \ref{sec:method}. The results of this work are presented in Section \ref{sec:results}, and are followed by a discussion in Section \ref{sec:dis}. The findings of our work are summarised in Section \ref{sec:concl}.

\section{Method}\label{sec:method}
\texttt{DustPy} \citep{Stammler2022}, based on the model by \cite{BirnstielT.2010}, is employed to simulate a protoplanetary disc, which includes viscous evolution of the gas, and coagulation, fragmentation, advection, and diffusion of the dust. \texttt{DustPy}  is coupled with \texttt{SyMBAp} (Lau \& Lee in prep.), which is a parallelized version of the symplectic direct $N$-body algorithm \texttt{SyMBA} \citep{Duncan1998}.

\subsection{Disk model}
\subsubsection{Gas component}
We considered a protoplanetary disc around a Solar-type star. The disc is assumed to be axisymmetric and in vertical hydrostatic equilibrium. The initial gas surface density $\Sigma_{\text{g,init}}$ is set according to the self-similar solution of a viscous disc by \cite{Lynden-Bell1974} such that
\begin{equation}
        \Sigma_{\text{g,init}} = \frac{M_\text{disk}}{2\pi r_\text{c}^2}\left( \frac{r}{r_\text{c}}\right) ^{-1}\exp\left(-\frac{r}{r_\text{c}}\right)
,\end{equation}
with the distance from the star $r$, the initial mass of the disc $M_\text{disk}$ , and the characteristic radius $r_\text{c}$. We set $M_\text{disk}=0.1M_\odot$ and $r_\text{c}=100\text{ au}$, which imply $\Sigma_{\text{g,init}}\approx 1400 \text{ g cm}^{-2}$ at $r=1$ au. The gas disc viscously evolves over time $t$ according to the advection--diffusion equation:
\begin{equation}
        \frac{\partial \Sigma_\text{g}}{\partial t}=\frac{3}{r}\frac{\partial}{\partial r}\left[ r^{1/2} \frac{\partial}{\partial r}(\nu \Sigma_\text{g} r^{1/2}) \right],
\end{equation}
where the back reaction from the dust is neglected. The \cite{Shakura1973} $\alpha$-parametrization is adopted for the kinematic viscosity $\nu$ such that
\begin{equation}
\nu = \alpha c_\text{s} H_\text{g}
,\end{equation}
where  $c_\text{s}$ is the speed of sound and $H_\text{g}$  the disc scale height. The viscosity parameter $\alpha=10^{-3}$ is set in this work. The disc scale height is defined by $H_\text{g}\equiv c_\text{s}/\Omega_\text{K}$, where the local Keplerian orbital frequency $\Omega_\text{K}=\sqrt{GM_\ast/r^{3}}$ with the gravitational constant $G$ and the mass of the central star $M_\ast$. The isothermal sound speed is used and given by $c_\text{s} = \sqrt{k_\text{B}T/\mu}$ with the Boltzmann constant $k_\text{B}$, the midplane temperature $T,$ and the mean molecular weight of the gas $\mu=2.3m_\text{p}$. The disc is assumed to be passively irradiated by the Solar luminosity, which gives a midplane temperature profile of
\begin{equation}
T\approx263.2 \left( \frac{r}{\text{au}}\right) ^{-1/2} \text{K}.
\end{equation}
This setup yields the dimensionless gas disc scale height
\begin{equation}
        \hat{h}_\text{g}\equiv\frac{H_\text{g}}{r}\approx0.0326\left( \frac{r}{\text{au}}\right)^{1/4}.
\end{equation}
The midplane pressure gradient parameter $\eta$ is then given by
\begin{equation}
        \eta=-\frac{\hat{h}_\text{g}^2}{2}\frac{\partial\ln P}{\partial\ln r} \label{eq:eta}
,\end{equation}
where $P$ is  the midplane gas pressure, which describes the degree of `sub-Keplerity' of the gas. A logarithmic radial grid is adopted that spans from 1 to 100 au with 99 cells for a disc gap
at 10 au, and from 10 to 100 au with 66 cells for a disc gap at 75 au. Additional grid refinement is imposed around the exterior pressure bump of the gap (see Section \ref{sec:gap}).

\subsubsection{Dust component}
The initial dust surface density $\Sigma_{\text{d,init}}$ is given by
\begin{equation}
        \Sigma_{\text{d,init}} = Z \ \Sigma_{\text{g,init}}
,\end{equation}
with the global dust-to-gas ratio $Z$ set at the Solar metallicity of 0.01. We follow the size distribution of the interstellar medium \citep{Mathis1977} for the initial dust grain sizes. The maximum initial size is set at $1\mu\text{m}$ and the internal density of $1.67\text{ g cm}^{-3}$ is assumed. A total of 141 dust mass bins are
used, logarithmically spaced from $10^{-12}$ to $10^8$ g. Each mass species is evolved in time through transport with the advection--diffusion equation \citep{Clarke1988} coupled to growth and fragmentation with the Smoluchowski equation. The fragmentation velocity is assumed to be $10\text{ m s}^{-1}$ and dust aggregates are assumed to fragment at collision velocities above this value. The Stokes number St$_i$ of each mass bin $i$ is then calculated by considering the
Epstein and the Stokes I regimes. The dust scale height of each mass species $H_{\text{d},i}$ is calculated according to \cite{Dubrulle1995}:
\begin{equation}
H_{\text{d},i}=H_\text{g}\sqrt{\frac{\alpha}{\alpha+\text{St}_i}}.\label{eq:h_d}
\end{equation}
Further details of the gas and dust evolution algorithm are described in \cite{Stammler2022}.

\subsection{Disk gap} \label{sec:gap}
A stationary Gaussian gap is created in the disc following the model in \cite{Dullemond2018}. A pre-existing planet in the gap is not considered in this work. As $\alpha \Sigma_{\text{g}}$ is a constant in our disc model at a steady state, the disc gap can be created from a smooth disc by adopting a modified $\alpha$-parameter with radial dependence $\alpha'(r)$ given by
\begin{equation}
        \alpha'(r) = \alpha/F(r),
\end{equation}
where the function
\begin{equation}
        F(r)=\exp\left[ -A \exp\left(-\frac{(r-r_0)^2}{2w^2} \right) \right] 
,\end{equation}
where $A$ is  the amplitude of the  gap, $r_0$ its  location, and $w$ its width. We fixed $A=2$ and tested two gap locations $r_0=\{10,75\}$ au. At $r_0=10$ au, $w$ is set to 2 au. At $r_0=75$ au, $w$ is scaled by $r^{13/16}$, yielding $w\approx10.3$ au. This is motivated by the radial scaling of the gap width given by the empirical formula provided by \cite{Kanagawa2017} for the structure of a planet-induced gap. Combined with the disc model, the prescribed disc gap also results in a local maximum of gas surface density at the outer edge, which is also a local maximum of pressure. We note that the sign of $\eta$ in Eq. \ref{eq:eta} changes across the pressure maximum, where $\eta>0$ implies sub-Keplerian gas and $\eta<0$ implies super-Keplerian gas.

\subsection{Planetesimal formation} \label{sec:SI}
\begin{figure}
        \centering
        {\graphicspath{{./fig/}} \input{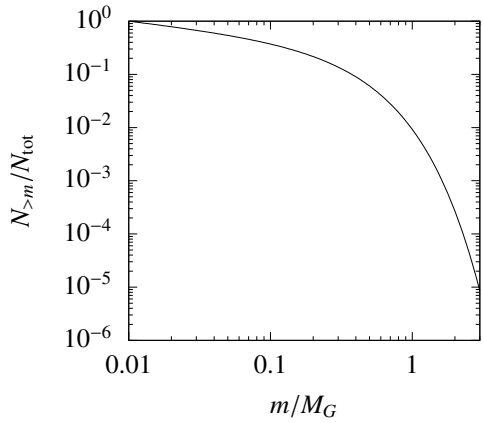}}
        \caption{Truncated power-law cumulative mass distribution adopted following \cite{Abod2019}.}
        \label{fig:Ntot}
\end{figure}
\begin{figure}
        \centering
        {\graphicspath{{./fig/}} \input{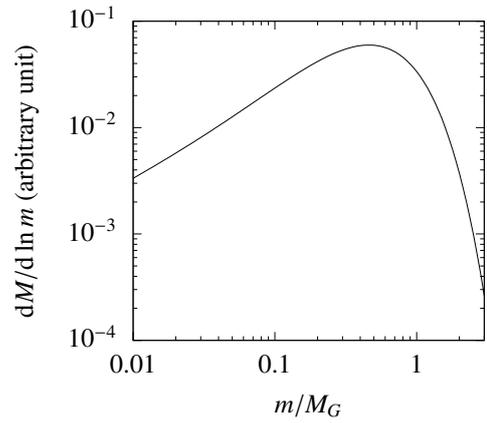}}
        \caption{Total mass in each logarithmic mass bin in the adopted planetesimal mass distribution. The mass in the most massive bin is about 17.9 times more massive than the bin at $10^{-2}M_G$.}
        \label{fig:dM}
\end{figure}

As dust accumulates at the exterior pressure bump of the disc gap, the dust is converted into planetesimals based on the prescriptions in \cite{DrazkowskaJ.2016} and \cite{SchoonenbergDjoeke2018} with the adoption of the smooth planetesimal formation activation function $\mathcal{P_\text{pf}}$ in \cite{Miller2021}. A smooth activation function that centres around a midplane dust-to-gas ratio $\rho_\text{d}/\rho_\text{g}$ of unity is expected to be more probable than a sharp activation of the streaming instability, because the latter causes planetesimals to form in discrete pulses with sharp temporal fluctuations in our tests. The smooth activation function is given by
\begin{equation}
        \mathcal{P_\text{pf}}=\frac{1}{2}\left[ 1+\tanh\left( \frac{\log(\rho_\text{d}/\rho_\text{g})}{n}\right)  \right]
,\end{equation}
with the smoothness parameter $n$ set to 0.03 in this work. $\mathcal{P_\text{pf}}$ is evaluated at each radial cell and the local dust surface density in each mass bin $i$ is removed by
\begin{equation}
        \frac{\partial \Sigma_{\text{d},i}}{\partial t}=-\mathcal{P_\text{pf}}\Sigma_{\text{d},i}\frac{\zeta}{t_{\text{set},i}}=-\mathcal{P_\text{pf}}\zeta\Sigma_{\text{d},i}\text{St}_i\Omega_\text{K}
,\end{equation}
where $\zeta$ is the planetesimal formation efficiency per settling time, which we set to 0.05 corresponding to a planetesimal conversion time of 40 settling times when the dust-to-gas ratio is unity. The settling time $t_{\text{set},i}$ of mass bin $i$ is given by
\begin{equation}
    t_{\text{set},i}=\frac{1}{\text{St}_i\Omega_\text{K}}.
\end{equation}
We note that $\zeta$ is not well constrained while the effect of using a different value is not studied in this work. The local dust surface density loss is then summed over all mass bins and added to the local planetesimal surface density $\Sigma_\text{plts}$, that is,
\begin{equation}
        \frac{\partial \Sigma_\text{plts}}{\partial t}=-\sum_i \frac{\partial \Sigma_{\text{d},i}}{\partial t}.
\end{equation}

Planetesimals are then realised from the radial profile of $\Sigma_\text{plts}(r)$. We adopted the cumulative mass distribution resulting from the fitting to the streaming instability simulations by \cite{Abod2019}. This has the form of an exponentially truncated power law such that the number fraction of planetesimals above mass $m$ is given by
\begin{equation}
        \frac{N_{>m}}{N_\text{tot}}=\left( \frac{m}{m_{\min}}\right) ^{-0.3}\exp\left( \frac{m_{\min}-m}{0.3M_G}\right), \label{eq:dN}
\end{equation}
for $m\geq m_{\min}$ with the minimum planetesimal mass $m_{\min}$ and the characteristic planetesimal mass $M_G$. The form of Eq. (\ref{eq:dN}) is shown in Fig. \ref{fig:Ntot}. \cite{SchaeferUrs2017} noted that $m_{\min}$ is sensitive to the spacial resolution of the streaming instability simulation. Nonetheless, the total mass in each mass bin in logarithmic scale can be estimated by
\begin{equation}
        \text{d} M \propto -\frac{\text{d}N_{>m}}{\text{d}m} m^2 \text{d}\ln m. \label{eq:dM}
\end{equation}
We set $m_{\min}=10^{-2}M_G$ in this work, where the peak of $\text{d} M$ is about 17.9 times that at $m_{\min}$  as shown in Fig. \ref{fig:dM}. As Eq. (\ref{eq:dN}) $\to 0$ only when $m\to \infty$, we artificially set the upper limit of $m$ at $3M_G$ in the algorithm of realisation, which implies that a small number fraction of planetesimals of about $8.48\times 10^{-6}$ is lost. \cite{Klahr2020} considered the critical mass for gravitational collapse of a dust clump in the presence of turbulent diffusion. We adopted this mass as $M_G$, which is given by
\begin{align}
M_G&=\frac{1}{9}\left( \frac{\delta}{\text{St}} \right)^{3/2} \hat{h}_\text{g}^{3} M_\odot  \label{eq:M_G}\\
&\approx7.22\times10^{-3}\left( \frac{\delta/\text{St}}{10^{-4}/10^{-2}} \right)^{3/2} \left( \frac{\hat{h}_\text{g}}{0.058} \right)^3M_\oplus\nonumber,
\end{align}
with the small-scale diffusion parameter $\delta,$ which is independent of $\alpha$ in our model. As noted by \cite{Johansen2006}, there is disagreement over the measurement of the relative strength of turbulent viscosity and turbulent diffusion in the literature, and this remains an active research topic \citep[e.g.][]{Schreiber2018}. We tested two values of $\delta=\{10^{-4},10^{-5}\}$. This is motivated by the measurements in streaming instability simulations by \cite{Schreiber2018}. Depending on the box sizes, the values of the measured small-scale diffusion parameter in the radial direction range from $10^{-6}$ to $10^{-4}$ for the midplane dust-to-gas ratio of 1. As $M_G\propto\delta^{3/2}$, a huge number of planetesimals are produced for the case with $\delta=10^{-6}$ in our preliminary runs, which is computationally unaffordable. Therefore, this value of $\delta$ is not tested in this work. The potential effect is discussed in Section \ref{sec:m_g}. Also, St is evaluated by taking the density-averaged value across all mass bins in the local radial cell. We note that \cite{Abod2019} also provided a characteristic mass based on the balance between the tidal force and self-gravity, which has a dependence of $\Sigma_{\text{g}}\Omega_\text{K}^{-4}$ and is independent of the local diffusivity. This would result in a characteristic mass of $\sim10M_\oplus$ at 80 au in our disc model, which is not physically probable.

In each simulation, the semi-major axis $a$ of a planetesimal is first randomly drawn using $\Sigma_\text{plts}(r)$ as a radial distribution function and $M_G$ is evaluated at the local radial cell. The planetesimal mass $m$ is then drawn from Eq. (\ref{eq:dN}). At each communication between \texttt{DustPy} and \texttt{SyMBAp}, if the total planetesimal mass accumulated in terms of surface density exceeds this value, a planetesimal is then realised and subtracted from the accumulated mass. The eccentricity $e$ is randomly drawn from a Rayleigh distribution with the scale parameter $10^{-6}$. The inclination $i$ in radians is also drawn from a Rayleigh distribution but with the scale parameter $5\times10^{-7}$. Other angles of the orbital elements are drawn randomly from 0 to $2\pi$. The physical radius $R_\text{p}$ is calculated by assuming an internal density $\rho_\text{s}$ of 1.5 $\text{g cm}^{-3}$. The planetesimal surface density is then subtracted according to the $m$ and $a$ of the realised planetesimal. If the total planetesimal mass in the local cell is not enough, the planetesimal mass from the neighbouring cells is used for the subtraction as well. Afterwards, another value of $m$ is drawn immediately and a planetesimal with mass $m$ is realised until the remaining accumulated planetesimal mass is less than $m$. The last drawn value of $m$ is retained for the next communication step such that the realisation is not biased towards lower mass. As this process does not guarantee that all mass in $\Sigma_\text{plts}$ can be realised, the residual is accumulated for the next communication step. Further details on the communication step are provided  in Section \ref{sec:num}.

\subsection{Planetesimal evolution} \label{sec:plts}
The realised planetesimals and a Solar-mass central star are then evolved by \texttt{SyMBAp} with full gravitational interactions as well as additional subroutines to include gas drag, type-I damping and migration, and pebble accretion. If two bodies collide, they are assumed to merge completely. Collisions are therefore perfectly inelastic in this work. At each communication step, other than the newly formed planetesimals in terms of surface density, the radial profiles of the gas component and the dust component are passed to \texttt{SyMBAp}. The gas component includes
\begin{itemize}
        \item the gas surface density $\Sigma_\text{g}$;
        \item the midplane temperature $T$;
        \item the gas disc scale height $H_\text{g}$;
        \item the midplane gas density $\rho_\text{g}$, and;
        \item the midplane pressure gradient parameter $\eta$,
\end{itemize}
and the dust component, for each mass bin $i$, includes
\begin{itemize}
        \item the Stokes number St$_i$;
        \item the dust disc scale height $H_{\text{d},i}$, and;
        \item the dust surface density $\Sigma_{\text{d},i}$.
\end{itemize}
The mass of the accreted pebbles is also passed to \texttt{DustPy} and subtracted from the dust surface density as further discussed in Section \ref{sec:pa}. As the planetesimals gradually gain mass, they are referred to as embryos, protoplanets, or planetary cores in this work. Generally, planetesimal refers to a body that has not gained significant mass, while embryo refers to a body that has grown by more than an order of magnitude. Protoplanet refers to a massive dominating body in the context of viscous stirring among a crowd of planetesimals, while planetary core refers to a body that is sufficiently massive and has the potential to accrete gas and form a giant planet. Nonetheless, these bodies are not strictly distinctive and have no fundamental difference in the simulations. They are all treated as fully interacting particles.

\subsubsection{Gas drag and type-I torques}
Planetesimals and planets experience the combined effect of aerodynamic gas drag and type-I torques due to the planet--disc interaction in this work. Generally, in typical disc environment at 10 au, gas drag ($\propto m^{-1/3}$) is more dominant for small bodies, while type-I torques ($\propto m$) gradually overtake for $m \gtrsim 10^{-5}M_\oplus$. As gas accretion and feedback are not considered, the transition to the high-mass regime (type-II) is not included.

We adopted the aerodynamic gas drag by \cite{Adachi1976} where
\begin{equation}
        \mbox{\boldmath $a$}_\text{drag}=-\left( \frac{3C_\text{D}\rho}{8R_\text{p}\rho_\text{s}}\right) v_\text{rel}\mbox{\boldmath $v$}_\text{rel},
\end{equation}
where $C_\text{D}$ is  the drag coefficient and $\mbox{\boldmath $v$}_\text{rel}$ is  the relative velocity between
the planetesimal and the gas. The gas flow is assumed to be laminar and cylindrical, where the magnitude is given by $v_\text{K}(1-|\eta|)$ with the local Keplerian velocity $v_\text{K}$. As the planetesimals in this work are much larger than a km in diameter, the large Reynolds number case is generally applicable for $C_\text{D}$, and we set the value to 0.5 \citep{Whipple1972}. The gas density $\rho$ at the position of the  planetesimal $z$ above the midplane is given by $\rho=\rho_\text{g}\exp(-0.5z^2/H_\text{g}^2)$.

For type-I damping and migration, we adopted the prescription based on dynamical friction by \cite{Ida2020}. The timescales for the isothermal case and finite $i$ while $i<\hat{h}_\text{g}$ (Appendix D of \cite{Ida2020}) are implemented. The evolution timescales of the semi-major axis, eccentricity, and inclination are defined respectively by
\begin{equation}
\tau_a\equiv-\frac{a}{\text{d}a/\text{d}t},\tau_e\equiv-\frac{e}{\text{d}e/\text{d}t}, \tau_i\equiv-\frac{i}{\text{d}i/\text{d}t}.
\end{equation}
These timescales are given by, with $\hat{e}\equiv e/\hat{h}_\text{g}$ and $\hat{i}\equiv i/\hat{h}_\text{g}$,
\begin{equation}
\tau_a = \frac{t_\text{wav}}{C_\text{T}\hat{h}_\text{g}^2}\left[ 1+\frac{C_\text{T}}{C_\text{M}}\sqrt{\hat{e}^2+\hat{i}^2}\right], \label{eq:tau_a}
\end{equation}
\begin{equation}
\tau_e =1.282t_\text{wav}\left[ 1+ \frac{(\hat{e}^2+\hat{i}^2)^{3/2}}{15} \right],
\end{equation}
\begin{equation}
\tau_i =1.805t_\text{wav}\left[ 1+ \frac{(\hat{e}^2+\hat{i}^2)^{3/2}}{21.5} \right].
\end{equation}
The characteristic time $t_\text{wav}$ \citep{Tanaka2002} is given by
\begin{equation}
        t_\text{wav}=\left( \frac{M_\ast}{m}\right) \left( \frac{M_\ast}{\Sigma_{\text{g}}r^2}\right) \left( \frac{\hat{h}_\text{g}^4}{\Omega_\text{K}}\right),
\end{equation}
where $\Sigma_{\text{g}}$ and $\hat{h}_\text{g}$ are retrieved from the local radial cell of the disc model. Due to frequent close encounters in the planetesimal belt, the semi-major axes may briefly depart greatly from the instantaneous locations, resulting in an unphysical $t_\text{wav}$, which is then evaluated at the instantaneous $r$ of the embryo instead of its semi-major axis, in contrast to the statement in \cite{Ida2020}. The coefficients $C_\text{M}$ and $C_\text{T}$ are given by
\begin{equation}
C_\text{M}=6(2p_\Sigma-q_T+2),  \label{C_M}
\end{equation}
\begin{equation}
C_\text{T}=0.73+1.08p_\Sigma+0.87q_T,\label{C_T}
\end{equation}
where $p_\Sigma\equiv-\text{d}\ln\Sigma_{\text{g}}/\text{d}\ln r$ and $q_T\equiv-\text{d}\ln T/\text{d}\ln r$, which are evaluated with the local radial cell as well as the immediate neighbouring ones. The three timescales are then applied to the equation of motion,
\begin{equation}
\mbox{\boldmath $a$}=-\frac{v_\text{K}}{2\tau_a}\mbox{\boldmath $e$}_\theta-\frac{v_r}{\tau_e}\mbox{\boldmath $e$}_r-\frac{v_\theta-v_\text{K}}{\tau_e}\mbox{\boldmath $e$}_\theta-\frac{v_z}{\tau_i}\mbox{\boldmath $e$}_z
,\end{equation}
in the cylindrical coordinate system using the notation $(r,\theta,z)$ where the velocity of the embryo $\mbox{\boldmath $v$}=(v_r,v_\theta,v_z)$. Also, $v_\text{K}$ is evaluated at the instantaneous $r$ of the embryo. \cite{Ida2020} noted that local uniformity on the scale of $H_\text{g}$ is assumed in the derivations. This condition is satisfied for the disc gap implemented in this work, as described in Section \ref{sec:gap}, where the half width is wider than the local $H_\text{g}$ at both locations of 10 and 75 au.

\subsubsection{Pebble accretion} \label{sec:pa}
The planetesimals are formed in a dust-enhanced location, where significant growth by pebble accretion is expected. We adopted the pebble accretion efficiency $\epsilon_\text{PA}$ by \cite{LiuBeibei2018} and \cite{OrmelChrisW2018}, which is defined as the mass fraction of the pebble flux accreted by the planetesimal or planet. This prescription includes both the ballistic regime and the settling regime of pebble accretion, and considers the local disc conditions and the orbit of the pebble-accreting embryo. In particular, the $e$ and $i$ of the embryo are taken into account when evaluating the relative velocity with respect to the pebble, which is critical in a planetesimal belt because viscous stirring is significant. As a substructured disc is considered in this work, the `pebble formation front' model by \cite{LambrechtsM.2014a} cannot be applied, as explicitly noted by the authors, where a finite and positive $\eta$ is assumed.

We note that $\eta$ changes sign across the pressure bump and requires careful treatment. For the relative velocity between the pebble and an embryo in circular orbit, the direction of the Hill shear also changes sign for super-Keplerian pebbles drifting outwards from the inner orbits. Therefore, the existing expression combining the headwind- and shear-dominated regimes is still valid (Eq. (10) in \cite{LiuBeibei2018}) and the absolute value of $\eta$ should be used in this case. As the radial profile of $\eta$ is given in a radial grid from \texttt{DustPy}, to capture the narrow region where $\eta$ can be very close to zero, the value of $\eta$ is interpolated at the radial position of each embryo before the absolute value is taken.

Following \citet{Drazkowska2021}, the pebble-accretion rate is calculated by summing the respective rates for each mass bin $i$ at the local radial cell, which is given by multiplying the corresponding $\epsilon_{\text{PA},i}$ to the pebble flux $\dot{M}_\text{peb}=2\pi rv_{\text{drift},i} \Sigma_{\text{d},i}$. The pebble drift velocity $v_{\text{drift},i}$ of mass bin $i$ is given by \citep{Weidenschilling1977}
\begin{equation}
        v_{\text{drift},i}=2\text{St}_i|\eta|r\Omega_\text{K}. \label{vdrift}
\end{equation}
The accretion rate can be summarised as
\begin{align}
        \dot{m}&=\sum_i \epsilon_{\text{PA},i} 2\pi r v_{\text{drift},i}\Sigma_{\text{d},i}\\
        &=\sum_i \epsilon_{\text{PA},i} 4\pi r^2 \text{St}_i|\eta| \Omega_\text{K}\Sigma_{\text{d},i}.
\end{align}
As $\eta$ is expected to cross zero in our disc model, we note that this means $v_\text{drift}=0$ when $\eta=0$ with Eq. (\ref{vdrift}) and $\epsilon_\text{PA}$ is undefined. In our implementation, this would give $\dot{m}=0,$ as the supply of pebbles by headwind is halted. The mass of the accreted pebbles of each mass bin in the respective radial cell is then subtracted from the dust component of the disc in the next immediate communication step. Nonetheless, the contribution from the turbulence of the gas and the diffusion of the dust on $v_\text{drift}$ are neglected, and only the pebbles supplied by the headwind is considered; that is, a more conservative pebble accretion rate is adopted. The possible implications of this are further discussed in Section \ref{sec:pb}.

As this work neglects the gas feedback from the embryos onto the disc, the simulation is stopped once any particle reaches one-tenth of the local pebble isolation mass $m_\text{iso}$, which is given by \citep{LambrechtsM.2014}:
\begin{equation}
        m_\text{iso}=20\left(\frac{\hat{h}_\text{g}}{0.05}\right)^3M_\oplus.
\end{equation}
Nevertheless, \citep{Sandor2021} reported that the pebble isolation mass may be about two to three times larger at a pressure bump, while the results presented in \cite{BitschBertram2018} do not support this conclusion. 

\subsection{Numerical setup}\label{sec:num}
The time step in \texttt{DustPy} $\Delta t_\text{disk}$ is variable and is determined by the rate of change of the gas and dust surface densities, while \texttt{SyMBAp} requires a fixed time step of $\Delta t_\text{nbod}$ for the symplecticity. For all the simulations in this work, $\Delta t_\text{disk}>\Delta t_\text{nbod}$. Therefore, after $\Delta t_\text{disk}$ is evaluated in \texttt{DustPy}, it is rounded down to the nearest integral multiple of $\Delta t_\text{nbod}$. With the time step determined, \texttt{DustPy} takes one step, and \texttt{SyMBAp} takes a number of $\Delta t_\text{disk}/\Delta t_\text{nbod}$ steps in parallel. Communication is then made via MPI, where the data of the newly formed planetesimals (Section \ref{sec:SI}), the relevant gas and dust components (Section \ref{sec:plts}), and the rounded $\Delta t_\text{disk}$ are passed from \texttt{DustPy} to \texttt{SyMBAp}, and, the data of the accreted pebbles (Section \ref{sec:pa}) are passed from \texttt{SyMBAp} to \texttt{DustPy}. Afterwards, \texttt{DustPy} and \texttt{SyMBAp} take their respective step(s) again in parallel and this process repeats until the simulation ends.

For the simulations with the disc gap at 10 au, $\Delta t_\text{nbod}=0.5$ yr is used and particles are removed if the heliocentric distance is less than 5 au or greater than $10^3$ au. Also, for the gap location of 75 au, $\Delta t_\text{nbod}=20$ yr is used and particles are removed if the heliocentric distance is less than 50 au instead. The additional subroutines for gas drag, type-I damping and migration, and pebble accretion are added to \texttt{SyMBAp} following the integration step in \cite{MatsumuraSoko2017} as
\begin{equation}
\mathcal{P}^{\tau/2}\mathcal{M}^{\tau/2}\mathcal{N}^{\tau}\mathcal{M}^{\tau/2}\mathcal{P}^{\tau/2},
\end{equation}
where the time step $\tau=\Delta t_\text{nbod}$, the operator $\mathcal{P}$ handles the effect of pebble accretion, $\mathcal{M}$ handles the effect of gas drag, type-I damping, and migration, and $\mathcal{N}$ is the second-order symplectic integrator in the original \texttt{SyMBAp}. Also, $\mathcal{P}$ and $\mathcal{M}$ operate in the heliocentric coordinates and $\mathcal{N}$ operates in the democratic heliocentric coordinates; therefore, coordinate transformations are done at each step.

In this work, two gap locations $r_0=\{10,75\}$ au and two values of the local diffusion parameter $\delta=\{10^{-4},10^{-5}\}$ are tested. Five simulations for each combination are conducted in order to minimise the statistical effect, which means a total of 20 simulations are conducted and presented in the following section.

\section{Results}\label{sec:results}
\begin{table*}
\caption{Summary of the combinations of parameters, the times required for each combination to reach the $t_0$ where the first planetesimal was formed, and the time when $0.1m_\text{iso}$ is reached. As the mass of the planetesimal is drawn randomly, there are some variations in the time required among the simulations with the same parameters. Further descriptions can be found in the respective subsections listed in the last column.}             
\label{table:param}
\centering          
\begin{tabular}{l l r r r }
\hline\hline
$r_0$ (au) & $\delta$ & Time to $t_0 \ (\times10^5\text{yr})$ & Time to $0.1m_\text{iso}$ $(t_0 + ... \times10^5\text{yr})$ & Section \\
\hline                    
   10 & $10^{-4}$ & $\{1.05, 1.11, 1.14, 1.14, 1.15\}$ & $\{0.34, 0.26, 0.22, 0.24, 0.24\}$ & \ref{sec:10.4}\\  
   10 & $10^{-5}$ & $\{0.93, 0.98, 1.03, 1.05, 1.06\}$ & $\{0.51, 0.46, 0.37, 0.34, 0.37\}$ & \ref{sec:10.5}\\
   75 & $10^{-4}$ & $\{4.9, 5.2, 5.3, 5.6, 6.0\}$ & $\{2.2, 1.9, 1.9, 1.4, 1.2\}$ & \ref{sec:75.4}\\
   75 & $10^{-5}$ & $\{4.5, 4.6, 4.6, 4.6, 5.1\}$ & $\{3.1, 2.9, 3.0, 3.0, 2.4\}$ & \ref{sec:75.5}\\
\hline                  
\end{tabular}
\end{table*}
In each simulation, some time is required for the gap to attain the prescribed form and accumulate enough dust to trigger planetesimal formation by streaming instability. We define the time at which the first planetesimal is realised as $t_0$ and Table \ref{table:param} summarises the time required to reach $t_0$ for each simulation. As the masses of the planetesimals are drawn randomly, simulations with the same parameter may not reach $t_0$ at the same time and a small variation is recorded. Simulations end when $0.1m_\text{iso}$ is reached and the respective times are also summarised in Table \ref{table:param}. Descriptions for each set of parameters are shown in the following subsections.

\subsection{Disk gap at 10 au and $\delta=10^{-4}$} \label{sec:10.4}
\begin{figure}
        \centering
        \includegraphics{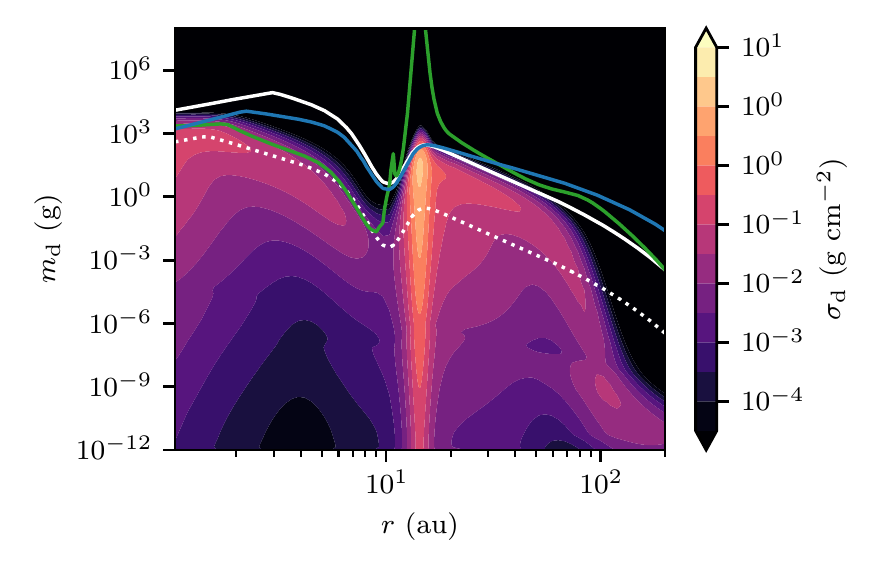}
        \caption{Dust distribution at $t_0$ of one of the five simulations for the disc gap at 10 au and $\delta=10^{-4}$. The heat map shows the radial profile of the midplane dust density $\sigma_\text{d}$ for different dust mass $m_\text{d}$. The white lines show the $m_\text{d}$ corresponding to $\text{St}=0.1$ (solid) and 0.01 (dotted), respectively. The green and blue lines show the drift and the fragmentation limits, respectively. The dust mass is shown to be limited by the fragmentation limit ($\gg$ mm) at the dust trap near 14 au.}
        \label{fig:disk}
\end{figure}
\begin{figure}
        \centering
        \includegraphics{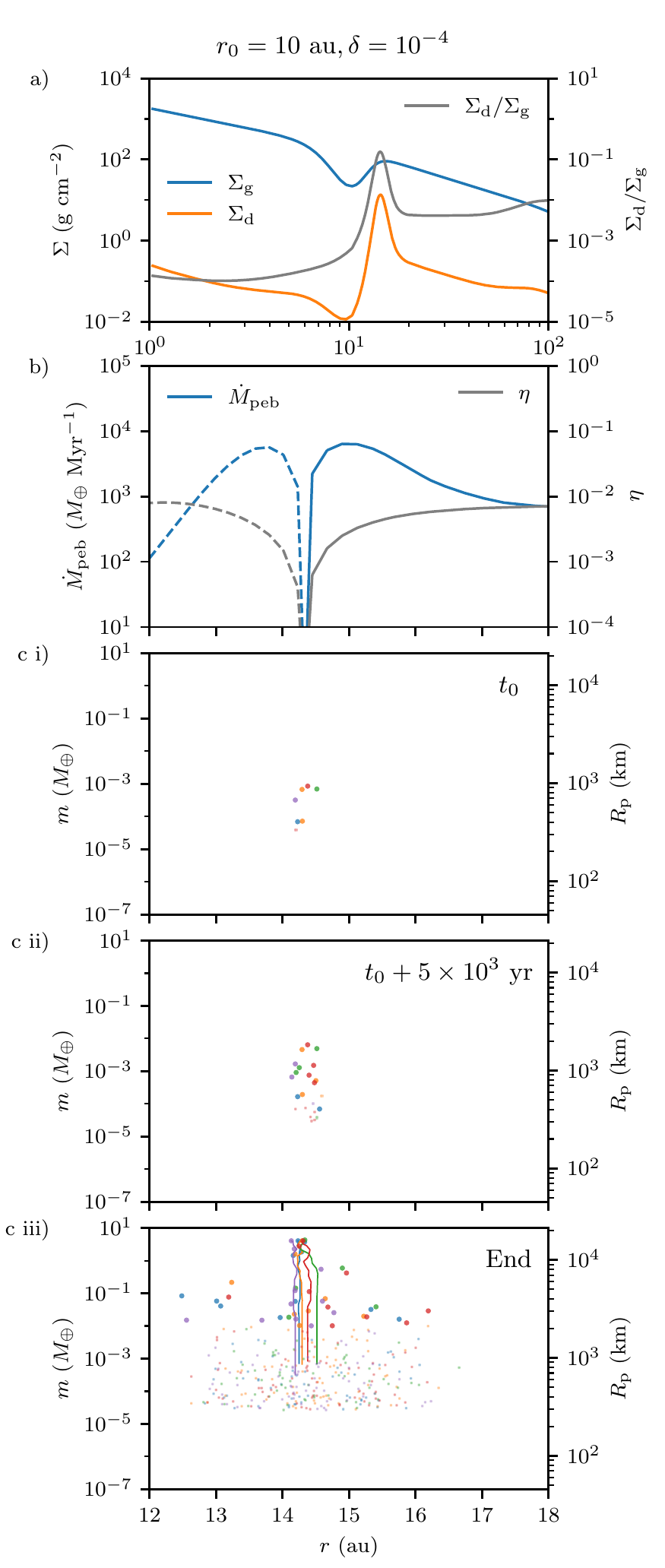}
        \caption{Simulation results for the disk gap at 10 au and $\delta=10^{-4}$, where $M_G\approx 3\times10^{-3}M_\oplus$. \textbf{(a)} Radial profiles in one of the five simulations of the dust and gas surface densities as well as the dust-to-gas ratio at $t_0$ when planetesimal formation starts. \textbf{(b)} Pebble flux $\dot{M}_\text{peb}$ and the pressure support parameter $\eta$ around the pressure bump from one of the five simulations, where the dashed lines denote negative values. \textbf{(c)} Time sequence of mass $m$ and semi-major axis $r$ of the planetesimals. Planetesimals that reach $10^{-2}M_\oplus$ by the end are denoted with large dots. Each colour shows one of the five simulations, which ended from $t_0+2.2\times10^4$ to $t_0+3.4\times10^4$ yr. Further descriptions are provided in the text.}
        \label{fig:4.10a}
\end{figure}
\begin{figure}
        \centering
        \includegraphics{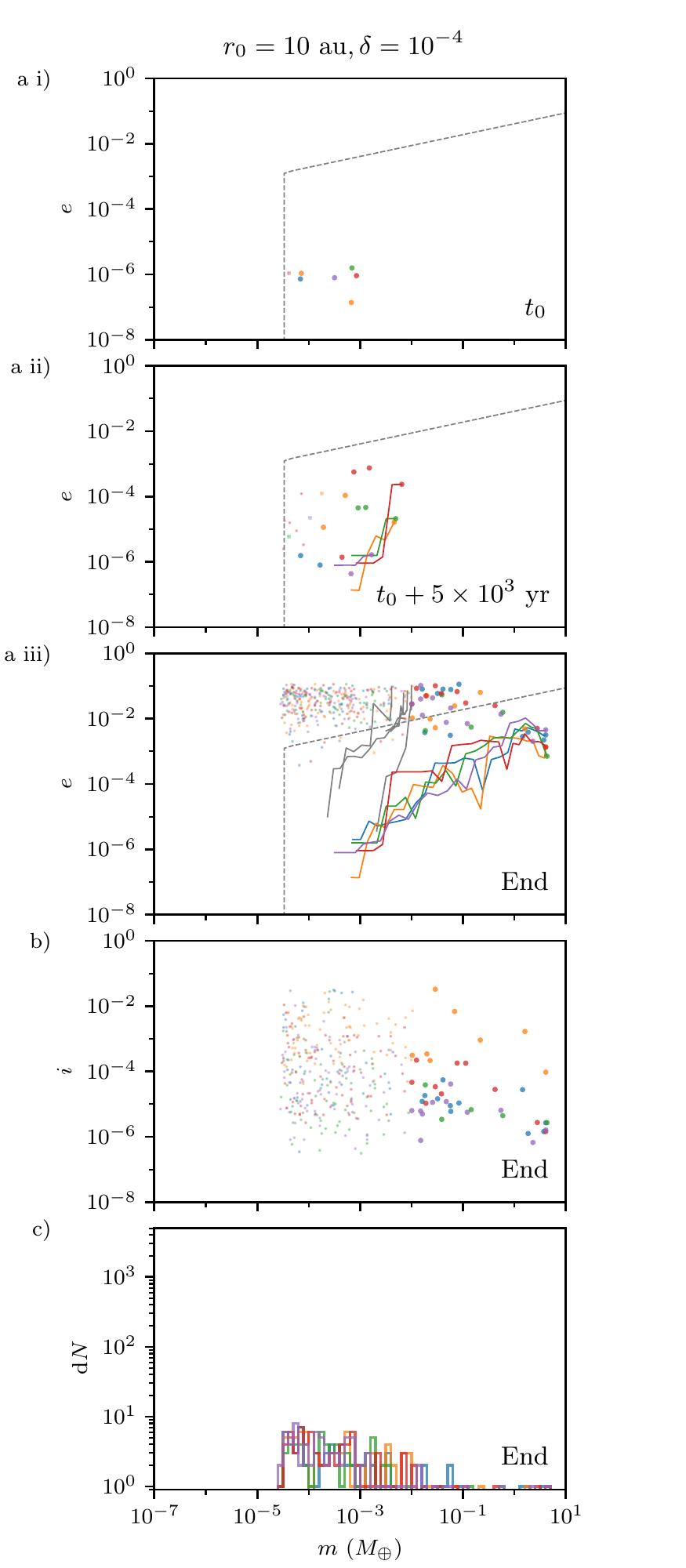}
        \caption{Further simulation results for the disc gap at 10 au and $\delta=10^{-4}$. \textbf{(a)} Time sequence of planetesimal eccentricity $e$ and mass $m$. The coloured trajectories follow the most massive bodies and the grey trajectories follow the most massive ones under $10^{-2}M_\oplus$ in each of the simulations. The dashed lines denote the pebble accretion onset mass discussed in Section \ref{vspa}. \textbf{(b)} Planetesimal inclination $i$ and mass $m$ at the end, where the $i$ of most bodies is still lower than the pebble disc scale height. \textbf{(c)} Differential mass distribution at the end of the model with ten mass bins in each decade. This shows that only a few massive cores are formed while the simulations stop before a large population of planetesimals is formed.}
        \label{fig:4.10b}
\end{figure}

Figure \ref{fig:disk} shows the radial dust distribution of one of the five simulations at $t_0$, where the heat map shows the profile of the midplane dust density $\sigma_\text{d}$ for different dust mass $m_\text{d}$. The drift and the fragmentation limits are shown by the green and blue lines, respectively. These diagrams  show that the dust mass is only limited by the fragmentation limit ($\gg$ mm) at the pressure bump near 14 au. The inward-drifting dust from the outer disc is trapped and coagulation continues. Planet formation by streaming instability occurs as enough dust accumulates at the pressure bump.

Figures \ref{fig:4.10a} and \ref{fig:4.10b} show the results of the five simulations for $r_0=10$ au and $\delta=10^{-4}$. Figure \ref{fig:4.10a}a shows the radial profiles of the gas and dust surface densities as well as that of the dust-to-gas ratio when planetesimals start to form in one of the simulations. We note that there is no significant difference in the disc profiles among the simulations and the disc profiles did not change drastically up to the end of the simulations. The disc gap is shown centred at 10 au as prescribed and dust accumulates at the outer edge of the gap. Across the local pressure maximum, as shown by the radial profiles from one of the simulations in Fig \ref{fig:4.10a}b, both $\dot{M}_\text{peb}$ and $\eta$ cross zero and switch sign, where the negative values are denoted by the dashed lines. However, just outside of this narrow region, the peaks of $\dot{M}_\text{peb}$ are almost up to $10^4 M_\oplus \ \text{Myr}^{-1}$ while $\eta$ is still lower than $10^{-2}$ at the these two locations.

In Fig. \ref{fig:4.10a}c, panels (i) to (iii) show the mass and semi-major axis of the planetesimals at different times. The colours show the results from each of the five simulations and the bodies that reached $10^{-2}M_\oplus$ are denoted with large dots. In this case, the characteristic planetesimal mass $M_G\approx 3\times10^{-3}M_\oplus$, while there are small variations in time and distance as the disc is slowly evolving and the planetesimals do not form at the same location $r$. We define $t_0$ as the time at which the formation of planetesimals has just started. At $t_0$ (Fig. \ref{fig:4.10a}c i), just enough dust has accumulated at the pressure bump and planetesimal formation starts around the local pressure maximum at about 14.2 au. Planetesimals continue to form and some grow rapidly through pebble accretion as in Fig. \ref{fig:4.10a}c ii at $t_0+5\times10^3$ yr. The five simulations ended from $t_0+2.2\times10^4$ to $t_0+3.4\times10^4$ yr as $0.1m_\text{iso}\approx4.1M_\oplus$ is reached in each of them (Fig. \ref{fig:4.10a}c iii). The presence of the narrow region with low pebble flux does not appear to have a significant effect on the growth of the planetesimals in this setup. The coloured lines in Fig. \ref{fig:4.10a}c (iii)\ show the trajectories of the most massive body in each simulation. The massive planetary cores remain near the pressure bump with slight inward migration, which also started scattering the smaller planetesimals.

Figure \ref{fig:4.10b} shows further details of the effect of viscous stirring. Panels (a) (i) to (iii) show the eccentricity and mass of the planetesimals at the respective times in Fig. \ref{fig:4.10a}c. In Fig. \ref{fig:4.10b}a, the trajectory of the most massive body at the end of each simulation is shown by the coloured line, and that for the most massive body with $m<10^{-2}M_\oplus$ is shown by the grey line. The dashed line shows a pebble accretion onset mass assuming $\eta=10^{-3}$ and St $=10^{-1}$, which is further discussed in Section \ref{vspa}. From Fig. \ref{fig:4.10b}a (i to ii), the planetesimals form early, and close to the massive end of the distribution they grow rapidly by pebble accretion and stir the planetesimals formed later on (referred to here as the `latecomers'). In Fig. \ref{fig:4.10b}a (iii), the most massive body in each simulation has relatively low eccentricity throughout the entire simulation. The latecomers, which are born after an embryo has already grown significantly to $\sim 1M_\oplus$, are being stirred to high eccentricity as shown by the grey trajectories. This halts pebble accretion, even though their inclinations are still lower than the pebble disc scale height as shown in Fig. \ref{fig:4.10b}b. Meanwhile, the role of gas drag does not appear to be significant in the results, which is likely because of the large ($\gtrsim$ 100km) size of the planetesimals in our model and the strong dynamical heating due to the rapidly formed massive cores. The role of viscous stirring and pebble accretion is further discussed in Section \ref{vspa}.

The differential mass distributions at the end (Fig. \ref{fig:4.10b}c) show that only a small number of massive cores ($\geq M_\oplus$) are formed in each of the simulations. In this setup, the simulations are quickly stopped as $0.1 m_\text{iso}$ is reached, which happens shortly after only a small number of planetesimals have formed. The form of the initial mass distribution is not clearly shown, although most planetesimals do not grow significantly. We note that less than ten mergers occur in each of the simulations, which have no significant effect on the growth and the final mass of the bodies.

\subsection{Disc gap at 10 au and $\delta=10^{-5}$} \label{sec:10.5}
\begin{figure}
        \centering
        \includegraphics{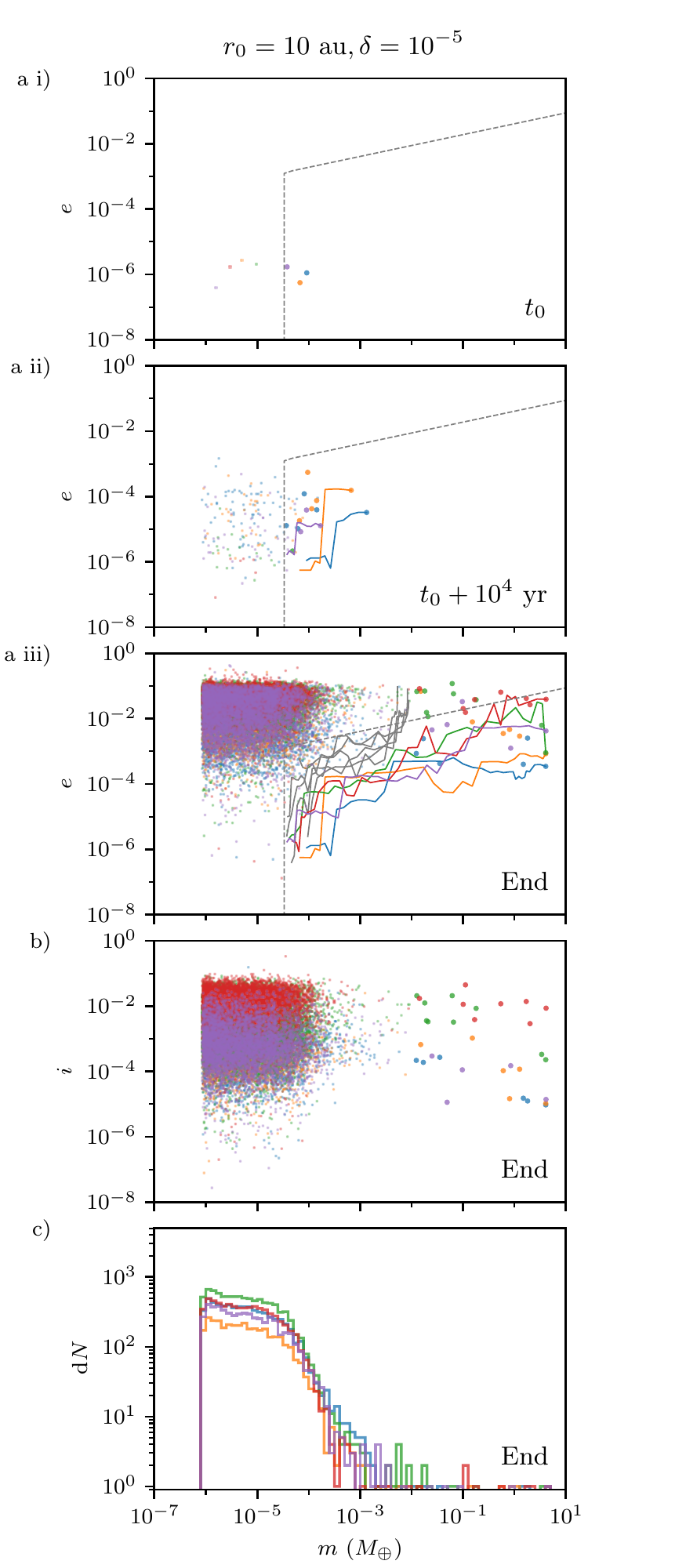}
        \caption{Simulation results for the disc gap at 10 au and $\delta=10^{-5}$, which ended from $t_0+3.4\times10^4$ to $t_0+5.1\times10^4$ yr. \textbf{(a)}  $e$-$m$ time sequence, showing that only the planetesimals formed early and relatively massive can accrete pebbles efficiently and stir up the latecomers. \textbf{(b)}  $i$-$m$ plot at the end of the simulations, showing that the inclinations of the small planetesimals are still not significantly larger than the pebble disc scale height. \textbf{(c)} Differential mass distribution at the end, showing that the majority of the planetesimals do not grow significantly by pebble accretion and retain the form of the initial mass distribution as shown in Fig. \ref{fig:Ntot}. Only a small number of massive cores are formed in the simulations.}
        \label{fig:5.10b}
\end{figure}
Another set of five simulations with $r_0=10$ au and $\delta=10^{-5}$ were conducted, for comparison with the results where $\delta=10^{-4}$. As suggested by Eq. (\ref{eq:M_G}), the change of $\delta$ means that $M_G$ is $10^{3/2}$, that is about 32 times lower than when  $\delta=10^{-4}$. With the characteristic mass $M_G\approx 9\times10^{-5}M_\oplus$, the number of planetesimals produced is much higher. The dust and gas surface densities, pebble flux, and $\eta$ show no significant difference with respect to the case of $\delta=10^{-4}$ as other parameters are unchanged. The five simulations ended from $t_0+3.4\times10^4$ to $t_0+5.1\times10^4$ yr.

Figure \ref{fig:5.10b} shows the results of models with $\delta=10^{-5}$. As previously, only the planetesimals that are formed early and massive can start pebble accretion immediately after formation. Similar to the case with $\delta=10^{-4}$, these bodies grow efficiently and stir up the population of smaller planetesimals. The grey trajectories in Fig. \ref{fig:5.10b}a (iii) show more clearly that some small bodies could still grow briefly by pebble accretion from about $10^{-4}M_\oplus$ to just below $10^{-2}M_\oplus$ but further growth is halted due to their high eccentricities. Although the inclinations of the small planetesimals shown in Fig. \ref{fig:5.10b}b are not much higher than the pebble disc scale height, they cannot still accrete pebbles. Figure \ref{fig:5.10b}c shows that the majority of the planetesimal population did not grow significantly. The form of the initial mass distribution (Fig. \ref{fig:Ntot}) is retained and only a small number of massive cores are formed. In each of the simulations, about 200 mergers occur, which is dominated by the massive cores accreting small planetesimals. The mass difference between the two populations is more than four orders of magnitude, and so these mergers have no significant effect on the final masses of the massive planetary cores.

\subsection{Disk gap at 75 au and $\delta=10^{-4}$} \label{sec:75.4}
\begin{figure}
        \centering
        \includegraphics{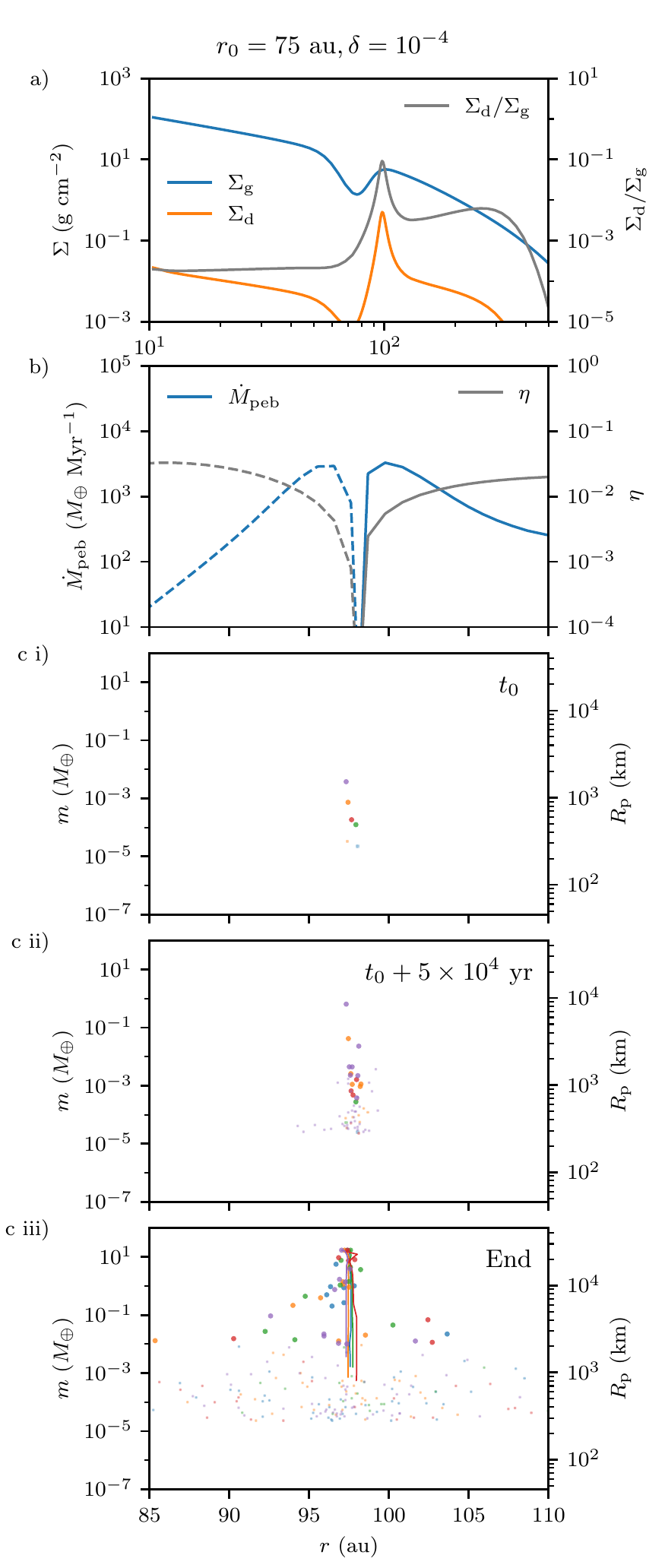}
        \caption{Simulation results for the disc gap at 75 au and $\delta=10^{-4}$. \textbf{(a)} The radial profiles of the disc when planetesimal formation starts. \textbf{(b)} Radial profiles of pebble flux $\dot{M}_\text{peb}$ showing that it is still high around the pressure bump, yet slightly lower compared to the models at 10 au shown in Fig. \ref{fig:4.10a}b. The pressure support parameter $\eta$ is generally a few times higher while the regions of low $\eta$ still coincide with the peaks of the pebble flux. \textbf{(c)}  $m$-$r$ time sequence, showing similar results except the planetesimals are formed later and the growth rate is slower. The five simulations ended from $t_0+1.2\times10^5$ to $t_0+2.2\times10^5$ yr.}
        \label{fig:4.75a}
\end{figure}
\begin{figure}
        \centering
        \includegraphics{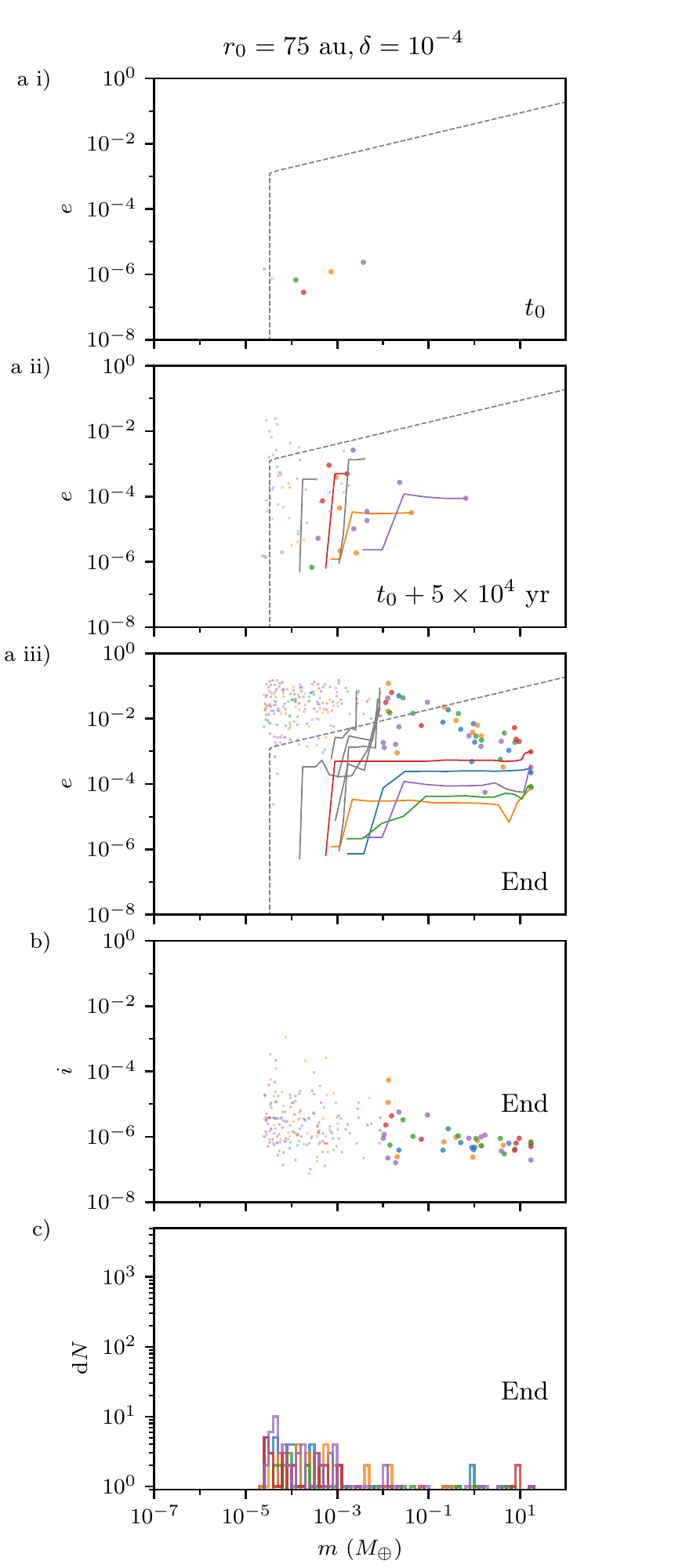}
        \caption{Further simulation results for the disc gap at 75 au and $\delta=10^{-4}$. \textbf{(a)}  $e$-$m$ time sequence, showing that a small number of massive cores are formed, which excite and prevent the growth of the latecomers. This is similar to the results to the case with $r_0=10$ au and $\delta=10^{-4}$. \textbf{(b)}  $i$-$m$ plot at the end of the simulations, showing that the inclinations remain very low for all bodies, i.e. around the values drawn at their formation. \textbf{(c)} Differential mass distribution at the end also shows only a small number of massive cores are formed and the simulations are stopped before a large population of planetesimals is produced. This is also similar to the case with $r_0=10$ au and $\delta=10^{-4}$.}
        \label{fig:4.75b}
\end{figure}
Figures \ref{fig:4.75a} and \ref{fig:4.75b} show the results of the five simulations for $r_0=75$ au and $\delta=10^{-4}$. Figure \ref{fig:4.75a}a shows the pressure bump centring at 75 au as prescribed and the outer pressure maxima is at about 100 au. Compared to the setup with the disc gap at 10 au, Fig. \ref{fig:4.75a}b shows that the pebble flux is only slightly lower near the pressure bump, while $\eta$ is a few times higher in general. The five simulations ended from $t_0+1.2\times10^5$ to $t_0+2.2\times10^5$ yr as $0.1m_\text{iso}\approx17.2M_\oplus$ is reached in each of them. Similarly, the $m$-$r$ time sequence (Fig. \ref{fig:4.75a}c) shows that the planetesimals formed early accrete pebbles efficiently and reach the pebble isolation mass in about 0.1 Myr. We note that $M_G$ is similar to that for the simulations with $r_0=10$ au, which is due the increase in $\hat{h}_\text{g}$ being mostly offset by the increase in the mass-averaged St.

Figure \ref{fig:4.75b} shows further details of the results, which are again similar those for the case where $r_0=10$ au and $\delta=10^{-4}$ (Fig. \ref{fig:4.10b}). In the time sequence (Fig. \ref{fig:4.75b}a), the planetesimals formed early grow by pebble accretion at a slower but still rapid rate and stir the latecomers to eccentric orbits that stop pebble accretion. Figure \ref{fig:4.75b}b shows that the inclinations of planetesimals remain even lower, which is around the values drawn at their realisation. Nonetheless, the small bodies still cannot accrete pebbles due to high relative pebble velocity and only a small number of massive bodies are formed, as shown in Fig. \ref{fig:4.75b}c.

\subsection{Disc gap at 75 au and $\delta=10^{-5}$} \label{sec:75.5}
\begin{figure}
        \centering
        \includegraphics{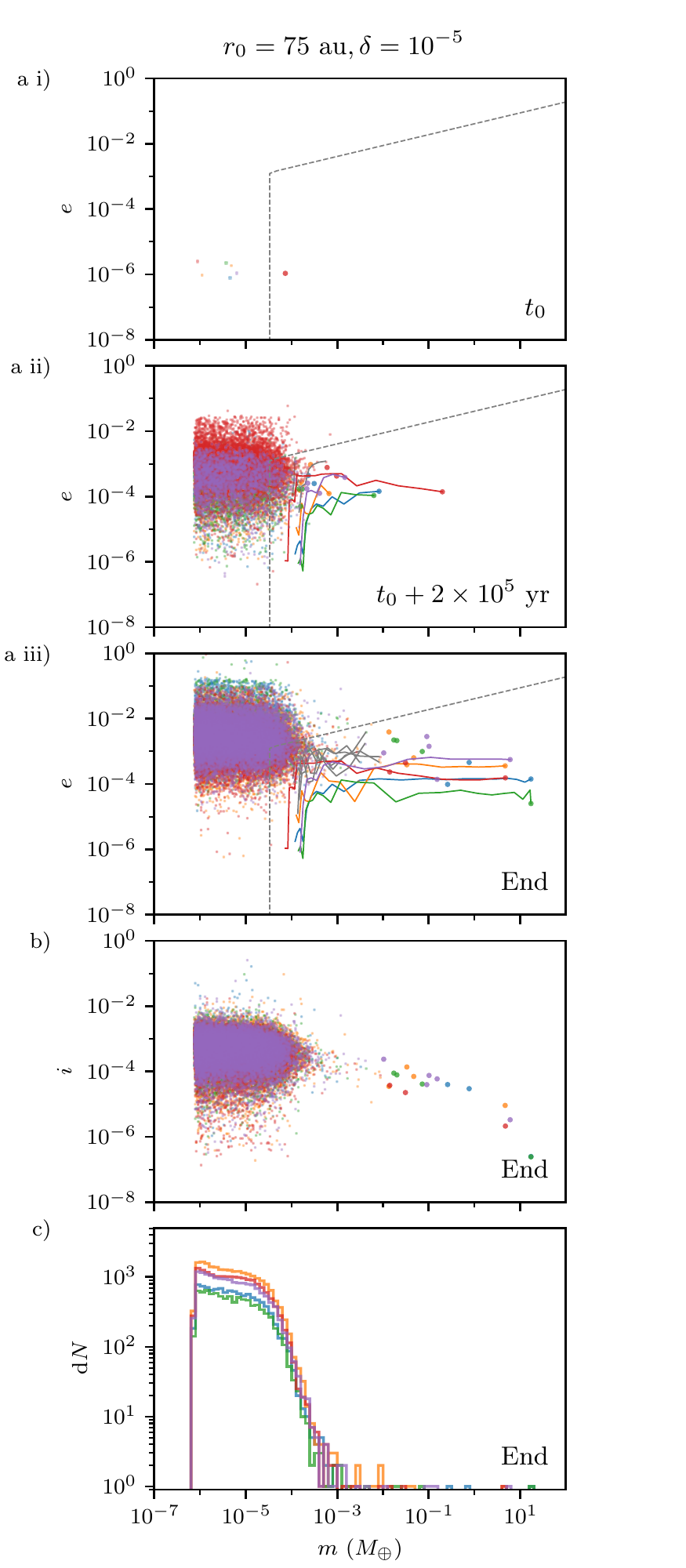}
        \caption{Simulation results for the disc gap at 75 au and $\delta=10^{-5}$, which ended from $t_0+2.4\times10^5$ to $t_0+3.1\times10^5$ yr. \textbf{(a)}  $e$-$m$ time sequence,
showing that a small number of massive embryos started growth early and stirred the majority of the late-forming planetesimals into eccentric orbits. \textbf{(b)}  $i$-$m$ plot at the end of the simulations, showing that the $i$ of all bodies remains far lower than the pebble disc scale height and confirms that the pebble relative velocity is critical. \textbf{(c)} Compared to the case of $r_0=10$ au and $\delta=10^{-5}$, the differential mass distribution shows an even smaller number of massive cores formed in these simulations.}
        \label{fig:5.75b}
\end{figure}
Figure \ref{fig:5.75b} shows the results of the five simulations for $r_0=75$ au and $\delta=10^{-5}$. The five simulations ended between $t_0+2.4\times10^5$ and $t_0+3.1\times10^5$ yr. Similar to the previous cases, only the planetesimals that formed early and relatively large can accrete pebbles efficiently, which is similar to the case where $r_0=10$ au and $\delta=10^{-5}$. The results also confirm that growth is dominated by the large embryos, which viscously stir the majority of the planetesimals. Pebble accretion again cannot operate for these excited bodies due to high pebble relative velocities, even though their inclinations are much lower than the pebble disc scale height. In comparison to the case where $r_0=10$ au and $\delta=10^{-5}$, a smaller numbers of larger embryos ($\geq M_\oplus$) are formed at the end.

\section{Discussions} \label{sec:dis}
\subsection{Pressure bump} \label{sec:pb}
\begin{figure}
        \centering
        \includegraphics{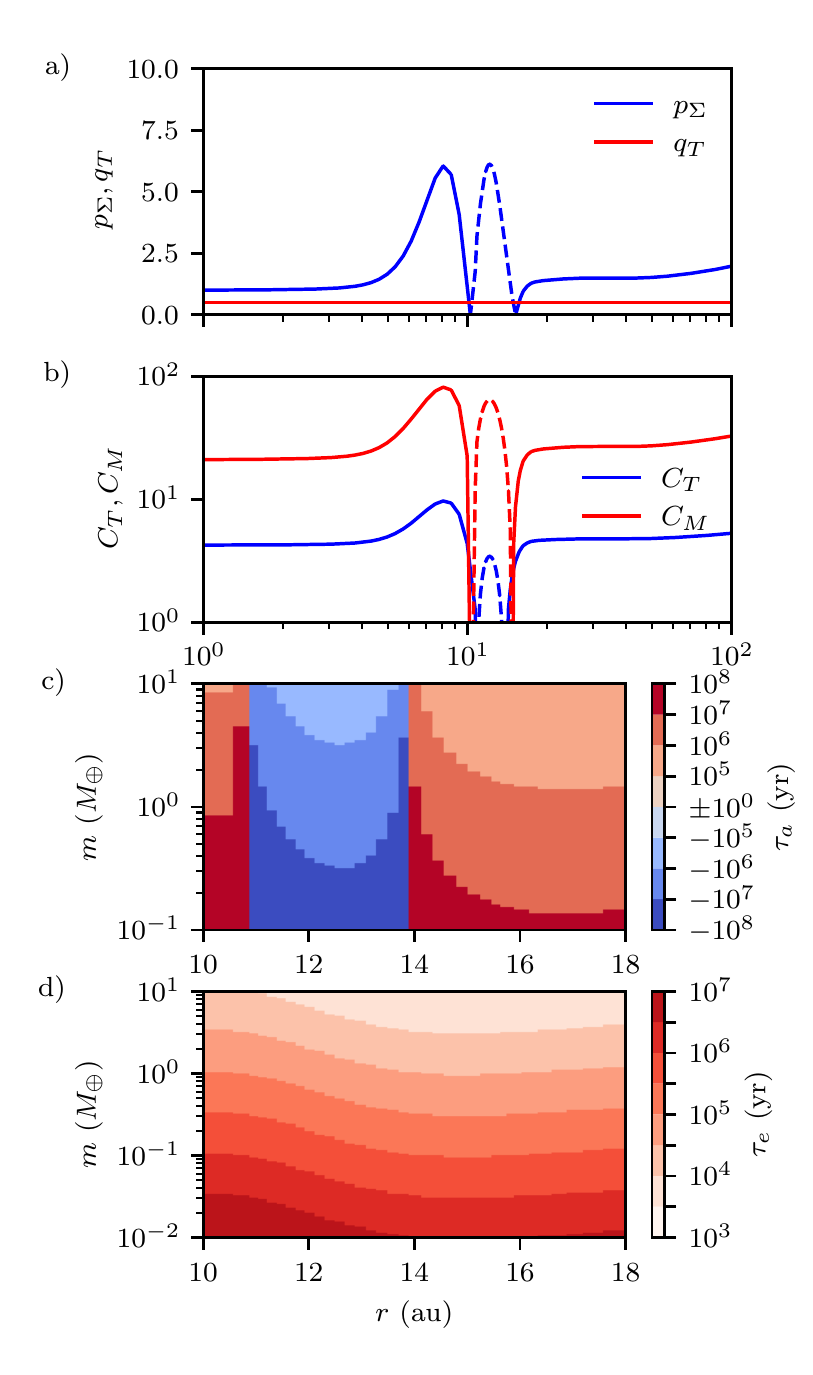}
        \caption{Radial profiles of the migration-related parameters at the end of one of the simulations with $r_0=10$ au. \textbf{(a)} $q_T$ remains constant in our disc model, while $p_\Sigma$ varies greatly and changes sign across the pressure bump. \textbf{(b)} Coefficients for $\tau_a$ also change sign but at slightly different locations according to  Eq. (\ref{C_M}) and (\ref{C_T}), respectively. \textbf{c)} $\tau_a$ near the pressure bump for embryos of different masses with $e=10^{-3}$ and $i=e/2$, where inward migration is denoted by red and outward migration is denoted by blue. The migration rate slows down and changes sign slightly interior ($\approx 14$ au) to the location of planetesimal formation ($\approx 14.2$ au). \textbf{(d)} $\tau_e$ at the same locations, which is $\propto t_\text{wav}\propto m^{-1}$. This shows type-I damping starts to become efficient when the embryos are $\gtrsim10^{-1}M_\oplus$.}
        \label{fig:mig}
\end{figure}
The results presented in this work show that the environment in a pressure bump is favourable to the rapid formation of massive planetary cores in numerous possible ways, which is in agreement with the results of \cite{Morbidelli2020}, \cite{GuileraOctavioMiguel2020} and \cite{Chambers2021}. Firstly, dust drifting from the outer disc is trapped and accumulated at the exterior edge of the disc gap. In this way, a pebble passing the planet orbit without being accreted is not lost to the inner disc, but can be accreted at later passages. This circumvents the issue of the requirement of large pebble masses \citep{Ormel2017}. Secondly, as the midplane dust-to-gas ratio gradually increases, it can reach the critical level for streaming instability. The grain sizes are also just limited by the fragmentation limit ($\gg \text{mm}$). Therefore, planetesimal formation by streaming instability is possible in this specific environment. The planetesimals are formed within the regions of high pebble density and can accrete them efficiently.

Furthermore, the headwind, which carries the dust or pebbles, is slower around the pressure bump. In the headwind regime, the relative velocity of the  pebbles is mostly determined by $\eta v_\text{K}$, which is applicable to a dynamically cold planetesimal belt. The pebble accretion onset mass in this case can be estimated using the following equation \citep{VisserRicoG.2016,Ormel2017}:
\begin{equation}
        M_\text{PA,hw}=\text{St}\ |\eta|^3M_\ast. \label{eq:mpahw}
\end{equation}
In both cases of $r_0=10$ au and 75 au, $\eta$ at the location of the peak of the pebble flux is a few times lower than that in a smooth disc. This greatly decreases the required mass for efficient pebble accretion, particularly in the outer disc as $\eta\propto\hat{h}_\text{g}^2$. Combining the decreased $M_\text{PA,hw}$ and the enhanced $\dot{M}_\text{peb}$, the planetesimals formed early that are well in the headwind regime can easily initiate rapid growth by pebble accretion. In our setup, the growth timescale in the settling regime at the pressure bump is $\sim 10^3-10^4$ yr for $r_0=10$ au and is a few times higher for $r_0=75$ au.

We note that $\dot{M}_\text{peb}$ is very low in the immediate vicinity of the peak of the pressure bump in our model, where the pebble drift velocity switches sign and crosses zero with $\eta$. This results from our assumptions that pebbles are only supplied by radial drift due to the headwind in the disc. In a more realistic scenario, pebbles are also transported by turbulent diffusion. This effect is negligible when the headwind dominates the supply of pebbles. However, at the pressure bump, the headwind changes direction and is weak within a narrow region (e.g. Fig. \ref{fig:4.10a}b and \ref{fig:4.75a}b). In this case, turbulent diffusion can supply pebbles to this region such that the pebble flux is always finite and smooth. Although these effects are modelled by \texttt{DustPy} in the disc, the prescription of the pebble accretion efficiency by \cite{LiuBeibei2018} and \cite{OrmelChrisW2018} is yielded from a model with a finite headwind, which ranges from $15$ to $60 \text{ms}^{-1}$. It remains uncertain whether or not the same prescriptions can be applied to pebbles transported by turbulent diffusion as well as those effectively transported by the relative radial motion of the embryo. Therefore, in this work, we only consider the pebble flux due to the headwind in the disc, which is proportional to $|\eta|$, which would result in a more conservative pebble accretion rate.

Nonetheless, the presence of this region of small $\dot{M}_\text{peb}$ does not appear to have a significant effect on the growth of the planetesimals, which is likely because of the finite width and the dynamical spreading of the planetesimal belt. The value of $\dot{M}_\text{peb}$ is also exceptionally high ($>10^3M_\oplus \ \text{Myr}^{-1}$) just outside of this region. However, the width of this low-pebble-flux zone shall change with the shape of the disc gap, which may have a more significant effect on the growth of the planetesimals with a different prescription.

In other works adopting a smooth disc model, as the embryos become massive, the rapid type-I migration is shown to cause these bodies to quickly move to the inner disc. However, in our model with the pressure bump, these massive planetary cores are retained, which is similar to the results for planet migration in a structured disc found by \cite{Coleman2016}. In the adopted migration prescription for an embryo with low eccentricity and inclination, the sign of $\tau_a$ is determined by the coefficient $C_\text{T}$ in Eq. (\ref{eq:tau_a}). Also, $C_\text{T}$ is given by Eq. (\ref{C_T}), which depends on $p_\Sigma$ and $q_T$. Figure \ref{fig:mig} shows the migration parameters at the end of one of the simulations with $r_0=10$ au as an example, where $\tau_a$ and $\tau_e$ are calculated assuming $e=10^{-3}$ and $i=e/2$. In our disc model, $q_T$ is constant throughout the disc, while $p_\Sigma$ varies greatly and switches sign across the pressure bump as shown in Fig. \ref{fig:mig}a. This implies that $C_\text{T}$ also switches sign and is negative slightly interior to the pressure maximum (Fig. \ref{fig:mig}b). As the migration direction changes from inward to outward going into the pressure bump from the exterior (Fig. \ref{fig:mig}c), this helps to retain the massive embryos at wide orbits as shown in our results in Section \ref{sec:results}. Nonetheless, we note that the mass dependence of the migration trap \citep[e.g.][]{Chrenko2022} is not studied in our work because gas accretion and gap opening, which occur upon the formation of massive planetary cores, are not considered in our model.

\subsection{Pebble accretion and viscous stirring}\label{vspa}
\begin{figure}
        \centering
        \includegraphics{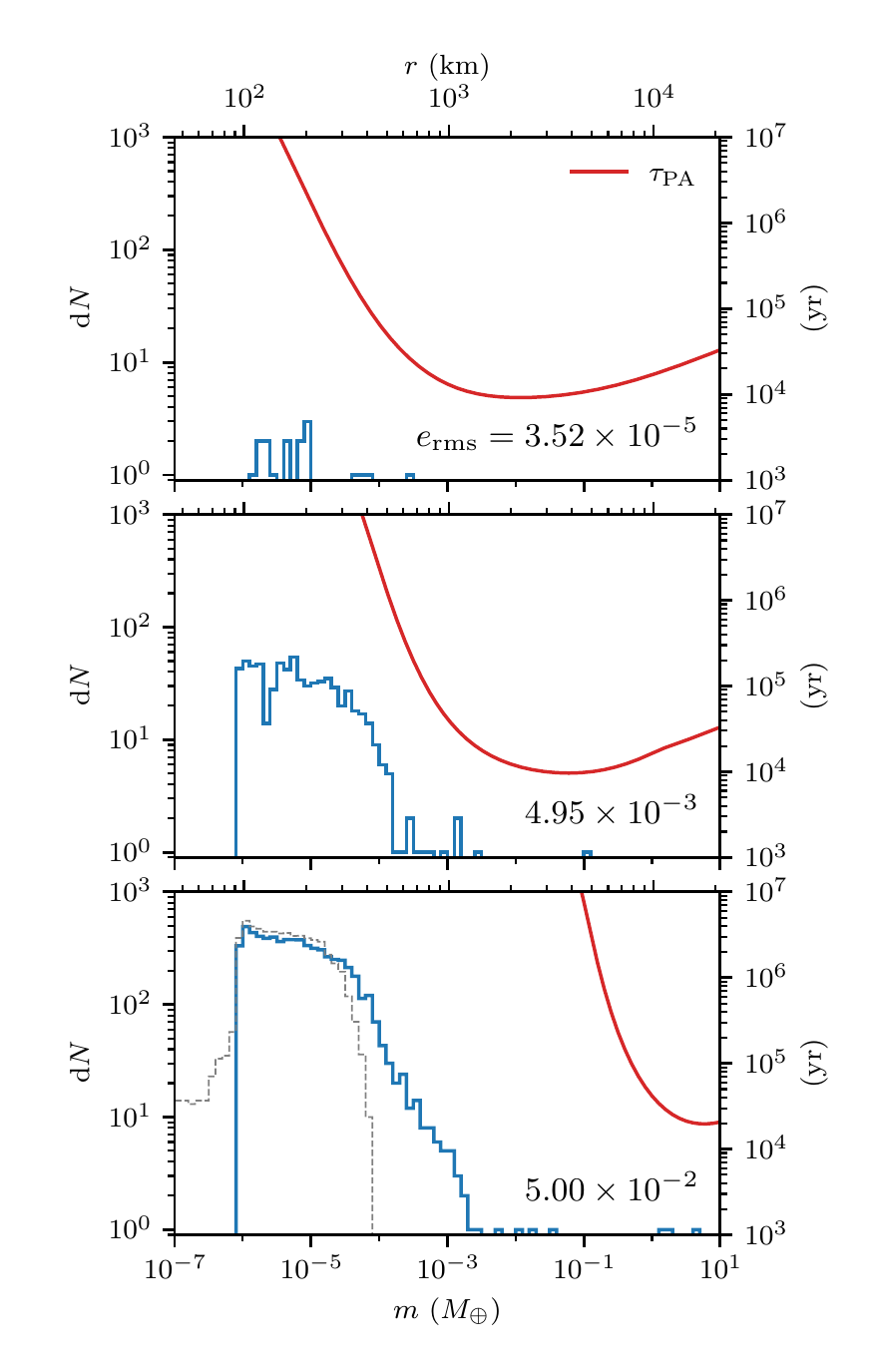}
        \caption{Differential mass distribution from one of the simulations with $r_0=10$ au and $\delta=10^{-5}$ at $t_0+5\times10^4$ yr (top), $t_0+2\times10^5$ yr (middle), and $t_0+3.7\times10^5$ yr (bottom) with ten mass bins in each decade. The red line in each panel shows the corresponding estimation of the pebble accretion timescale $\tau_\text{PA}$ with respect to mass for the instantaneous $e_\text{rms}$ of all bodies. The grey histogram in the bottom panel shows the initial mass of all planetesimals produced throughout the simulation. The evolution shows a window period exists where significant growth by pebble accretion is possible. This period starts from the time of the formation of the first planetesimal with $m>M_\text{PA,hw}$ and ends at the time when it becomes massive enough to excite the less massive bodies to high eccentricity in a short timescale.}
        \label{fig:hist}
\end{figure}
When a planetesimal is in an eccentric orbit, the relative velocity of the  pebbles is no longer dominated by the headwind but by the orbital velocity of the  planetesimal instead. In the adopted pebble accretion prescription provided by \cite{LiuBeibei2018} and \cite{OrmelChrisW2018}, the azimuthal pebble relative velocity is given by
\begin{equation}
        \Delta v_y = \max(v_\text{cir},v_\text{ecc}),\label{eq:dv}
\end{equation}
where $v_\text{cir}$ combines the headwind and the Hill regimes in the circular limit. Also, $v_\text{ecc}$ corresponds to the eccentric limit where $v_\text{ecc}=0.76ev_\text{K}$. For small planetesimals, pebble accretion mostly does not operate in the Hill regime. Therefore, the pebble accretion onset mass in Eq. (\ref{eq:mpahw}) can be refined into a more general form, where
\begin{equation}
        M_\text{PA}=\text{St}\ [\max(|\eta|,0.76e)]^3M_\ast.
\end{equation}
This mass is denoted by the dashed line in the $e$-$m$ plots in Section \ref{sec:results}, assuming $\text{St}=0.1$ and $\eta=10^{-3}$.

The shear-dominated (low-energy) regime of viscous stirring \citep{Ida1993} is applicable as the $e_m$ and $i_m$ of the newly born latecomers are very low before reaching the equilibrium values. The stirring timescale for eccentricity is much shorter than that for inclination and dynamical friction is ineffective \citep{Ida1990}. This is more significant in the results for $\delta=10^{-4}$ , where $e\ll i$ in general, even at the end of the simulations. Furthermore, in the results for $\delta=10^{-5}$, the excitation in $e$ occurs much earlier than that for $i$, while the equipartition of energy is only reached in some of the simulations. In this regime, the protoplanet--planetesimal viscous stirring timescale for eccentricity is given by \citep{Ida1993}
\begin{equation}
    \tau_{\text{vs},e}^{M-m}=7.20\times10^3\left( \frac{M}{M_\oplus}\right)^{-1/2}\left(\frac{a}{10\text{ au}}\right)^{3/2} \left(\frac{e_{m,\text{rms}}}{5\times10^{-3}}\right)^3\text{yr},
\end{equation}
where $e_{m,\text{rms}}$ is the root-mean-square value of $e_m$ and the mass of the protoplanet $M$. When a massive embryo has formed, the latecomers are stirred to high eccentricities within a very short timescale. Any newly formed planetesimals are excited to high eccentricity quickly, which results in a $M_\text{PA}\gg M_\text{PA,hw}$ for them. As a result, pebble accretion cannot operate even if the latecomers have mass greater than $M_\text{PA,hw}$.

Meanwhile, the $e$ of the massive bodies is damped efficiently by the type-I torques as $\tau_e\propto t_\text{wav}\propto m^{-1}$, which enables continuous pebble accretion. In the example for $r_0=10$ au as shown in Fig. \ref{fig:mig}d, type-I damping gradually becomes significant ($\lesssim10^6$ yr) for $m\gtrsim10^{-1}M_\oplus$. Thus, the latecomers are prevented from pebble accretion as a result of both viscous stirring from the massive planet formed earlier and the inefficient type-I damping.

This also implies that a window period exists between the time of the formation of the first planetesimal with $m>M_\text{PA,hw}$ and the time when it becomes massive enough to halt pebble accretion for the less massive bodies. During this window period, a few of the planetesimals formed with $m>M_\text{PA}\sim M_\text{PA,hw}$ can still grow by a few orders of magnitude in mass as shown in our results.

For example, Fig. \ref{fig:hist} shows the evolution of the differential mass distribution of one of the simulations with $r_0=10$ au and $\delta=10^{-5}$. The red line in each panel shows the corresponding pebble accretion timescale $\tau_\text{PA}\equiv m/\dot{m}$ with respect to $m$. $\tau_\text{PA}$ is estimated near the centre of the planetesimal belt at 14.2 au using the root-mean-square eccentricity of all bodies $e_\text{rms}$. The estimation also assumes $\dot{M}_\text{peb}=10^3M_\oplus \ \text{Myr}^{-1}$ and $\text{St}=0.1$. The top panel, at $t_0+5\times10^4$ yr, shows that when a massive body has not formed and $e_\text{rms}$ is low, $M_\text{PA}=M_\text{PA,hw}$ is about $10^{-4}M_\oplus$ where $\tau_\text{PA}\sim10^4$ yr. Rapid growth is possible for the bodies at the massive end of the planetesimal mass distribution. As a few embryos grow significantly in mass and excite $e_\text{rms}$ to above $0.76\eta$, the eccentric limit of pebble accretion becomes applicable. The middle panel, at $t_0+2\times10^5$ yr, shows that $M_\text{PA}$ starts to shift away from the majority of the planetesimals. The leading bodies dominate growth and the bottom panel, at $t_0+3.7\times10^5$ yr, shows that the small planetesimals are further excited and $M_\text{PA}$ increases drastically. The grey histogram shows the distribution of the initial mass of all planetesimals produced throughout the simulation. The bodies of $\gtrsim 10^{-3}M_\oplus$ were born within the window period such that the masses had grown by at least an order of magnitude before $M_\text{PA}$ overtook them.

The exact number of massive cores formed by the end of the simulations is then also determined by the rate of planetesimal formation, which is not explored in this work. As the transition between the headwind and the eccentric limits becomes critical in this scenario, the expression for $\Delta v_y $ in Eq. (\ref{eq:dv}) would benefit from further refinement, for a smoother and more physical expression.

We note that this mechanism is conceptually different from the `viscous stirring pebble accretion' in \cite{Levison2015}, where pebble accretion for smaller bodies is stalled due to high inclinations. In our model, the pebble disc scale height for each dust species is given by Eq. (\ref{eq:h_d}), which is about $0.32H_\text{g}$ for $\alpha=10^{-3}$ and $\text{St}=0.1$. Furthermore, in the results presented in Section \ref{sec:results}, the inclinations of the small bodies have not been significantly stirred away from the pebble disc in general but pebble accretion is already effectively stopped by the increased relative velocity of the  pebbles.

\subsection{Characteristic mass $M_G$}\label{sec:m_g}
The characteristic mass of the initial planetesimal mass distribution given by Eq. (\ref{eq:M_G}) is sensitive to the small-scale diffusion parameter where $M_G\propto\delta^{3/2}$. We only tested the values of $\delta=\{10^{-4},10^{-5}\}$ due to computational limits, while the measurements by \cite{Schreiber2018} range from $10^{-4}$ to $10^{-6}$. A strong dependence on the simulation domain size is also shown in the work of these latter authors.

As planetesimal accretion is inefficient in our parameter space, pebble accretion is only possible for the planetesimals with sufficient inital mass. For the results of $\delta=10^{-4}$ shown in Section \ref{sec:10.4} and \ref{sec:75.4}, $M_G>M_\text{PA,hw}$. In this case, planetesimals that are formed slightly later than the first one can still accrete pebbles and grow by at least an order of magnitude in mass. Also, $0.1m_\text{iso}$ is reached only after a small population of planetesimals have formed. The form of the initial mass distribution cannot be retrieved from the end results. In contrast, for $\delta=10^{-5}$ shown in Section \ref{sec:10.4} and \ref{sec:75.4}, $M_G\sim M_\text{PA,hw}$. Fewer bodies are formed with mass between the most massive cores and the vast population of small planetesimals. The distinction between these two classes is much clearer. As a result, the initial distribution of  planetesimals is mostly presented and preserved in this case.

Although the rapid formation of massive planetary cores is possible with the parameters tested, we expect that for a very low $\delta$, pebble accretion may not occur as $M_G\ll M_\text{PA,hw}$. Also, the timescale of runaway planetesimal accretion is given by \citep{Ormel2010}
\begin{align}
        \tau_{\text{rg}} = \ & 5.45\times10^7\left( \frac{m}{10^{-5}M_\oplus}\right)^{1/3}\left( \frac{M_\ast}{M_\odot}\right)^{-1/2}\left(\frac{\Sigma_\text{plts}}{10\text{ g cm}^{-2}}\right)^{-1} \times \nonumber\\
        &\left(\frac{\rho_\text{s}}{1.5\text{ g cm}^{-3}}\right)^{2/3}\left(\frac{a}{100\text{ au}}\right)^{3/2} \text{yr}.
\end{align}
Planetesimal accretion is also unlikely to be efficient enough to reach $M_\text{PA,hw}$ within the typical disc lifetime, particularly at the outer disc.
This will likely result in a population of planetesimals where the initial mass distribution is preserved. This consequence is also consistent with the recent work by \cite{Lorek2022}, who find that the seeds for pebble accretion are unlikely to form through planetesimal accretion beyond 5-10 au within the typical disc lifetime.

\subsection{Potential effects of a planet in the disc gap}
In this work, a planet in the disc gap is not considered while the results show that the growth of planetesimals by pebble accretion is also sensitive to viscous stirring. This implies that if there exists a planet in the gap, the planetesimal belt formed at the pressure bump may be heated and prevented from growth by pebble accretion. For instance, the fitting of a planet-induced gap in the simulations by \cite{Kanagawa2017} has a half width $\propto m^2$. Meanwhile, the half width of the heated zone of a planet \citep{Ida1993} scales with its Hill radius, where $\propto m^{1/3}$. This implies that there may exist a regime transition as the gap-opening planet grows in mass, because the gap expands faster than the heated zone. However, the exact width of the heated zone also depends on the $e$ and $i$ of the planetesimals as shown in \cite{Ida1993}. This requires further modelling, with the inclusion of gas accretion and feedback onto the gas disc. For instance, our model neglects the effect of the recently proposed thermal torque mainly due to heat released from solid accretion \citep{Benitez-Llambay2015,Masset2017,Guilera2021} and the effect of the dust torque \citep{Benitez-Llambay2018}.

\subsection{Other recent works}
Planet growth by pebble accretion in a substructured disc was also recently studied by \cite{Morbidelli2020}. In this work, the dust ring is assumed to be in a steady-state, which is described by a Gaussian distribution. The work of this latter author shows that planets may migrate slightly inwards but out of the dust-concentrated region, where growth by pebble accretion is significantly slowed down. In comparison, the dust-to-gas ratio in our model is much higher at the outer edge of the disc gap (e.g. Fig. \ref{fig:4.10a}a and \ref{fig:4.75a}a). As the disc gap is only prescribed towards the gas component, the dust component evolves around it and results in a much higher dust density at the pressure bump. Also, \cite{Morbidelli2020} adopted a migration prescription for a non-isothermal disc, in contrast to our model. As the coefficients of the migration torque are different between the isothermal and non-isothermal cases, the planet trap is located at a different position. Further investigations into the effect of migration in the context of substructured discs and planetary growth are required.

In the work by \cite{GuileraOctavioMiguel2020}, planet formation in an ice-line($\sim 3$ au)-induced pressure bump is also studied with a model that includes dust growth, planetesimal formation by streaming instability, pebble accretion, and planet migration. In this model, a uniform initial radius of 100 km for the planetesimals is adopted. When an embryo reaches lunar mass, the effects of pebble accretion, gas accretion, and planet migration are enabled for that object. Rapid formation of massive planetary cores is also demonstrated and the pressure bump acts as a planet trap for a planet with mass $\lesssim 10M_\oplus$. In comparison, we adopted a distribution of initial planetesimal mass and demonstrated the effect of viscous stirring on  pebble accretion using full $N$-body simulations. However, the initial planetesimal mass adopted in our work is generally much larger. The effect of planetesimal accretion is not noticeable because of the greater distance from the star. Once a massive core is formed, the neighbouring planetesimals are likely scattered and a new gap should form due to the planet--disc interactions; this new gap shall form a pressure bump at a new location.

\cite{Chambers2021} studied the formation of cold Jupiters in a substructured disc starting with embryos of $\sim 10^{-4} M_\oplus$. A series of eight pressure bumps is prescribed and the embryos are placed at the respective locations of the `pebble trap'. In addition to pebble accretion, gas drag, type-I torques, gas accretion, and gap opening are also included in this latter model. The results of these latter authors also show that massive planetary cores can form rapidly in a pressure bump, which acts as a migration trap as well. Meanwhile, the production of planetesimals is not considered in this latter model, something which is also favoured in the pressure bump, and pebbles are added to the disc locally after 400 orbital periods. As planetesimals are formed in a region of  concentrated dust, those above $M_\text{PA}$ can immediately start pebble accretion as shown in our model.

In the recent work by \cite{Jiang2022}, planetesimal growth in a clumpy ring and a pressure-induced dust ring are studied. In their case, a constant mass flux is supplied from the outer disc and a single Stokes number is assumed for the pebbles. In contrast, in the present work, we model the dust and gas evolution of the entire protoplanetary disc where mass is conserved globally. We also include the variation in the Stokes numbers of different mass species at different locations, which is critical to the initial planetesimal mass distribution and the pebble accretion efficiency.

A similar pebble-accretion efficiency prescription is adopted in the work of \cite{Jiang2022}. We note that the factor of $1/\eta$ from $\epsilon_\text{PA}$ is eliminated for the pebble-accretion rate while $\eta=0$ at the local pressure maximum. This implicitly assumes that the pebble flux due to the radial motion of the planet, turbulence in the disc, and diffusion of the dust is equivalent to that due to the headwind. In Section \ref{sec:pb}, we discuss the fact that the prescription of $\epsilon_\text{PA}$ is taken from models with non-zero headwind. Therefore, we have taken a more conservative approach by considering only the pebble flux due to the headwind for accretion in our work.

A migration prescription for a smooth disc with an artificial migration strength prefactor is also adopted in the work of these latter authors. Meanwhile, in the prescription by \cite{Ida2020} adopted in this work, the local slopes of the surface density and temperature profiles are taken into account, which naturally stop inward migration slightly interior to but not exactly at the pressure bump. Nonetheless, we note that these prescriptions assume a local uniformity on the scale of $H_\text{g}$. Further studies are required for a sharp pressure bump.

The work of \cite{Jiang2022} also shows that pebble accretion is efficient with low eccentricity, which is consistent with our results for the massive planetary cores formed. In addition, with a mass spectrum of planetesimals, the smaller or later-forming planetesimals  are significantly  excited and pebble accretion is halted as demonstrated in this work.

\cite{Jang2022} also studied the subsequent evolution of planetesimals formed by streaming instability in a smooth disc with a constant pebble flux. These authors modelled a planetesimal belt with a width of $\Delta r=\eta r$  at different locations of the disc. Their results show that rapid growth by pebble accretion is not possible at the outer disc due to the strong headwind, which drastically increases the pebble accretion onset mass. In the results with type-I migration, \cite{Jang2022} also find rapid inward migration for the embryos that reach $\sim1M_\oplus$. In contrast, as discussed in Section \ref{sec:pb}, planetesimals can still accrete pebbles efficiently in our results due to weakened headwind in the pressure bump, and are also retained near the pressure bump due to the change in the slope of the gas surface density.

\section{Conclusions} \label{sec:concl}
This work demonstrates the possibility of rapid formation of massive planetary cores in a pressure bump starting from interstellar-medium-sized dust. The dust and gas evolution code \texttt{DustPy} is used to model a protoplanetary disc consistently. \texttt{DustPy} is coupled with the parallelized $N$-body code \texttt{SyMBAp} to integrate a large number of planetesimals. According to the evolving disc, the planetesimals also experience the effects of gas drag, type-I migration, and pebble accretion.

As the micron-sized dust particles coagulate up to the centimetre to metre regime, the pressure bump traps the dust drifting from the outer disc. The locally enhanced dust-to-gas ratio can then trigger planetesimal formation by streaming instability. These km-sized planetesimals are naturally born in a location where the pebble flux is exceptionally large and the headwind is weakened. This allows the planetesimals that are formed both early and relatively massive to grow efficiently by pebble accretion even in the outer disc. Only a small number of massive cores ($\sim M_\oplus$) are formed as the later-formed planetesimals are excited into eccentric orbits, where pebble accretion is halted. The massive embryos are retained near the dust trap as the direction of migration switches to outward migration slightly interior to the local pressure maximum. A natural continuation of this work is the inclusion of gas accretion and feedback onto the protoplanetary disc in the model. This shall provide further insights into the open questions regarding the architecture of the Solar System as well as those of other exoplanetary systems.

Nonetheless, the characteristic masses of the planetesimals adopted in this work are limited by computational costs. As pebble accretion is unlikely to be efficient for even smaller planetesimals, further analysis of the initial mass distribution of planetesimals resulting from streaming instability is merited, which has been an active research topic; see for example \cite{Simon2016}, \cite{Simon2017}, \cite{SchaeferUrs2017}, \cite{Abod2019}, \cite{Li2019} and \cite{Rucska2020}.

As pebble accretion may not operate in a heated planetesimal belt due to the high relative velocity of the  pebbles, a pre-existing planet in the gap may immediately excite the newly formed planetesimals. This depends on the width of the gap and that of the heated zone. Neither a planet-induced gap nor the shape of the disc gap are explored in this work. These ingredients of our model require further investigation.

\begin{acknowledgements}
We thank Chris Ormel and Beibei Liu for the insightful discussions. We acknowledge funding from the European Research Council (ERC) under the European Union's Horizon 2020 research and innovation programme under grant agreement No 714769 and funding by the Deutsche Forschungsgemeinschaft (DFG, German Research Foundation) under grant 325594231 and Germany's Excellence Strategy - EXC-2094 - 390783311 and EXC 2181/1 - 390900948 (the Heidelberg STRUCTURES Excellence Cluster). JD was funded by the European Union under the European Union’s Horizon Europe Research \& Innovation Programme 101040037 (PLANETOIDS). Views and opinions expressed are however those of the authors only and do not necessarily reflect those of the European Union or the European Research Council. Neither the European Union nor the granting authority can be held responsible for them.
\end{acknowledgements}

\bibliographystyle{aa}
\bibliography{ms}

\end{document}